\documentclass[a4paper,11pt]{article}
\pdfoutput=1 

\usepackage{jcappub} 

\usepackage[T1]{fontenc} 
\usepackage{graphicx}
\usepackage{epsfig}
\usepackage{rotate}
\usepackage{amsmath}
\usepackage{amssymb}
\usepackage{amsfonts}
\usepackage{bm}
\usepackage{tablefootnote}
\usepackage{enumerate}
\usepackage{afterpage}
\usepackage{xcolor}
\usepackage{natbib, hyperref}

\UseRawInputEncoding

\newcommand{\ud}{\mathrm{d}}

\newcommand{\cH}{\mathcal{H}}
\newcommand{\fnl}{f_{\rm NL}}
\newcommand{\sfnl}{\sigma(f_{\rm NL})}

\def\be{\begin{equation}}
\def\ee{\end{equation}}
\def\bea{\begin{eqnarray}}
\def\eea{\end{eqnarray}}

\title{Multi-tracing the primordial Universe  with future  surveys}

\author{Mponeng Kopana$^1$, Sheean Jolicoeur$^{1,2}$, Roy Maartens$^{1,3,4}$}
\affiliation{$^{1}$Department of Physics \& Astronomy, University of the Western Cape, Cape Town 7535, South Africa\\
$^{2}$Department of Physics, Stellenbosch University, Matieland 7602, South Africa \\
$^{3}$Institute of Cosmology \& Gravitation, University of Portsmouth, Portsmouth PO1 3FX, United Kingdom\\
$^4$National Institute for Theoretical \& Computational Sciences (NITheCS), Cape Town 7535, South Africa }

\abstract{
The fluctuations generated by Inflation are  {nearly} Gaussian in the simplest models, but may be non-Gaussian in more complex models,  {potentially leading to signatures in the late Universe. In particular, local-type primordial non-Gaussianity  induces scale-dependent bias in tracers of the matter distribution. This non-Gaussian  imprint in the tracer power spectrum survives at late times on ultra-large scales where nonlinearity is negligible.} In order to combat the problem of growing cosmic variance on these scales, we  use a multi-tracer analysis that combines different tracers to maximise any primordial signal.  {Previous work has investigated the combination of a spectroscopic galaxy survey with a 21cm intensity mapping survey in single-dish mode.
We extend this work by considering instead the case where the 21cm intensity mapping survey is optimised for interferometer mode. As examples, we use two multi-tracer pairs of surveys:}  one at high redshift ($1\le z\le2$) and one at very high redshift ($2\le z\le5$). The 21cm surveys are  idealised surveys based on HIRAX and PUMA. We implement foreground-avoidance filters and use detailed models of the interferometer thermal noise. The galaxy surveys are idealised surveys based on Euclid and MegaMapper. Via a simple Fisher forecast, we illustrate the potential of the multi-tracer. Our results show a  {$\sim 20-30\%$} improvement in precision on local primordial non-Gaussianity from the multi-tracer.  {Furthermore, we investigate the effects on constraints of varying the parameter of non-Gaussian galaxy assembly bias and of varying the parameters of the intensity mapping foreground filters.}
We find that the non-Gaussian galaxy assembly bias parameter causes a greater change in the constraints on local primordial non-Gaussianity than the foreground filter parameters.}

\begin{document}
\maketitle
\date{\today}
\flushbottom

\section{Introduction}

Galaxy surveys in combination with cosmic microwave background (CMB) surveys  {and supernova Ia obervations} have laid the foundations for precision cosmology. The next generation of surveys will deliver even higher precision by covering larger sky areas and  deeper redshifts with better sensitivity.
Each of these surveys is destined to make major advances in the precision and accuracy of the cosmological parameters of the standard LCDM model. At the same time, future surveys will look for signals beyond LCDM. One such signal arises if there is significant non-Gaussianity in the primordial perturbations that seed the CMB fluctuations and the  large-scale structure.

Primordial non-Gaussianity (PNG) is a key probe of Inflation, which is currently the best framework that we have for the generation of seed fluctuations. The local type of PNG, described by the parameter $\fnl$, if found to be nonzero, will rule out the simplest Inflation models \cite{Achucarro:2022qrl}. Furthermore, if $0<|\fnl|\lesssim 1$, then many other  models can also be ruled out \cite{dePutter:2016trg,Meerburg:2019qqi}. In order to achieve this, a precision of $\sfnl<1$ is required.

The best current 1$\sigma$  constraint comes from the {\em Planck}  survey, via the CMB bispectrum \cite{Planck:2019kim}:
\be
f_{\rm{NL}} = -0.9 \pm 5.1\qquad  {(68\%~{\rm CL})}\,.
\ee
The CMB power spectrum does not carry an imprint of local PNG, and the same is true for the matter power spectrum. However, the power spectrum of a biased tracer, such as galaxies, does contain a signal of local PNG on ultra-large scales -- due to the phenomenon of scale-dependent bias \cite{Dalal:2007cu,Matarrese:2008nc,Slosar:2008hx}. Constraints from future CMB surveys will not be able to reach $\sfnl<1$ due to cosmic variance. In the case of galaxy surveys, the current best constraints are $\sfnl\gtrsim 20$  {(e.g. \cite{Rezaie:2021voi,mueller:2021jbt})}. Future galaxy surveys will improve on this, potentially reaching the CMB level of precision and beyond \cite{Raccanelli:2015vla,Alonso:2015uua,SKA:2018ckk,Krolewski:2023duv}. Although galaxy power spectrum constraints on $\fnl$ will be a powerful independent check on the CMB results, single-tracer constraints are unlikely to achieve 
$\sfnl<1$, again because of cosmic variance, which is highest on precisely the scales where the local PNG signal is strongest.

Since the local PNG signature in large-scale structure depends on the particular tracer, it is possible to improve the precision on $\fnl$ by combining the power spectra of different tracers. This multi-tracer approach 
can effectively remove cosmic variance  
\citep{Seljak:2008xr,McDonald:2008sh,Hamaus:2011dq,Abramo:2013awa}. As a result, the multi-tracer of future galaxy surveys with  intensity mapping of the 21cm emission of neutral hydrogen (HI IM surveys), can deliver significant improvements in the precision on $\fnl$ (e.g. \citep{SKA:2018ckk,Gomes:2019ejy,Jolicoeur:2023tcu}) and in some cases it can in principle reach $\sfnl<1$ (e.g. \cite{Alonso:2015sfa,Fonseca:2015laa,Ballardini:2019wxj,Squarotti:2023nzy,dAssignies:2023oxn}). The multi-tracer can  also mitigate some of the observational systematics in single-tracer surveys. 

 {Radio-optical  combinations have been intensively investigated (e.g. \cite{Alonso:2015sfa,Fonseca:2015laa,SKA:2018ckk,Gomes:2019ejy,Ballardini:2019wxj,Viljoen:2021ypp, Jolicoeur:2023tcu,Squarotti:2023nzy}). They are  well suited to the multi-tracer, since they combine surveys with very different clustering biases and systematics. Some of these investigations consider radio continuum surveys and photometric galaxy surveys.
Here we follow \cite{Fonseca:2015laa,Viljoen:2021ypp, Jolicoeur:2023tcu,Squarotti:2023nzy} in considering pairs made of a spectroscopic galaxy survey  and an HI intensity mapping survey. Previous papers have all used HI intensity mapping in single-dish mode. Here we use HI IM surveys in interferometer mode.  Surveys in this mode have different limiting scales, foreground effects and thermal noise, compared to single-dish mode surveys. Single-dish mode surveys probe larger scales than those in interferometer mode and are thus expected to provide better precision on $\fnl$. Nevertheless, it is interesting to see what is achievable in interferometer-mode surveys and whether it is worthwhile to combine them with galaxy surveys. In addition, an interferometer-mode survey on the proposed PUMA array \cite{PUMA:2019jwd} reaches higher redshifts than any proposed single-dish survey. Combining a PUMA-like survey with a galaxy survey like MegaMapper  \cite{Schlegel:2022vrv}  enables us to investigate the highest (post-reionisation) redshift multi-tracer pair of an HI IM and a galaxy survey.} 

We use nominal simplified surveys since we perform   Fisher forecasting which does not take into account most systematics. Our main aim is to demonstrate the advantage of the multi-tracer at high to very high redshifts with combined galaxy--HI probes, rather than to make forecasts for specific surveys.

 The paper is structured as follows. In section \ref{Multi-tracer power spectra}, we provide a brief demonstration of multi-tracer power spectrum of two tracers.  In section \ref{Surveys}, we presented the characteristics of the galaxy and HI intensity mapping surveys we are using. We discuss the survey's instrumentation and brief noise requirements in section \ref{Noise}. The avoidance of foreground contamination for HI intensity mapping is presented in section \ref{HI IM  foreground avoidance}. We also discuss the limited range of scales needed to obtain the cosmological information from both spectroscopic and radio surveys in section \ref{Maximum and minimum scales}. The Fisher forecast is discussed in Section \ref{Fisher forecast}, along with how we combined the data from the two surveys. Section \ref{Conclusion} provides a summary of our main conclusions.


\section{Multi-tracer power spectra}\label{Multi-tracer power spectra}
At linear order in perturbations, the Fourier number density or brightness temperature contrast of tracer $A$ is  
\begin{equation}
\delta_A(z, \bm{k}) = \Big[b_A(z)+f(z)\mu^{2}\Big]\delta(z, \bm{k})\;,  \label{e2.1}
\end{equation}
where $\delta$ is the  matter density contrast, $b_A$ is the (Gaussian) clustering bias for the tracer  $A$, $\mu = \bm{\hat{k}} \cdot \bm{{n}}$ is the projection along the line-of-sight direction $\bm n$, and  {$f = -\ud\ln\delta/\ud\ln(1+z) $ is the linear growth rate in LCDM}. We use a single-parameter model for the biases \cite{Agarwal:2020lov}, 
\begin{equation}
b_{A}(z) = b_{A0}\, {\alpha_A}(z)\;,\label{e2.2}
\end{equation}
where we assume that  the redshift evolution $ {\alpha_A}(z)$ is known. The power spectra for 2  tracers are given by 
\begin{equation}
\big \langle \delta_A(z,\bm{k})\delta_{B}(z,\bm{k}') \big \rangle = (2\pi)^{3}P_{AB}(z,\bm{k})\delta^{\mathrm{D}}(\bm{k}+\bm{k}')\;, \label{e2.3}
\end{equation}
so that
\begin{equation}
P_{AB}=\big(b_A+f\mu^2 \big)\big(b_B+f\mu^2 \big)P
\;,\label{e2.4}  
\end{equation}
where $P$ is the matter power spectrum (computed with CLASS \citep{Blas:2011rf}). When $B=A$ we use the shorthand $P_A \equiv P_{AA}$.

In the presence of local PNG, the  bias acquires a scale-dependent correction: 
\begin{equation}
\hat b_A(z,k) = b_{A}(z) + b_{A\phi}(z)\frac{f_{\rm NL}}{\mathcal{M}(z,k)}\,, \quad \mathcal{M}(z,k) = \frac{2}{3\Omega_{m0}H_{0}^{2}}\frac{D(z)}{g_{\rm in}}T(k)\,k^{2}\;, \label{e2.5}
\end{equation}
where $T(k)$ is the matter transfer function (normalized to 1 on very large scales), $D$ is the growth factor (normalized to 1 today) and $g_{\rm in}=(1+z_{\rm in})D_{\rm in}$ is the initial growth suppression factor, with $z_{\rm in}$ deep in the matter era. For $\Lambda$CDM, $g_{\rm in} = 1.274$. 
The modelling of the non-Gaussian bias factor $b_{A\phi}$ is in principle determined from halo simulations, but currently it remains uncertain \citep{Barreira:2020kvh,Barreira:2021dpt,Barreira:2022sey,Barreira:2023rxn, Fondi:2023egm,Adame:2023nsx}. A simplified model is given by 
\begin{equation}
b_{A\phi}(z) = 2 \delta_{\rm c}\big[b_{A}(z)-p_A\big]\,,\label{e2.6}
\end{equation}
where  the critical collapse density is given by $\delta_{\rm c} =1.686\,$ and $p_A$ is a  {tracer-dependent constant in the primordial non-Gaussian bias $b_{A\phi}$, arising from assembly bias}. For the simplest case of a universal halo model, $p_A=1$ for all tracers $A$ \cite{Dalal:2007cu,Matarrese:2008nc}. 

 {For our fiducial model,} we use \eqref{e2.6} with the following $p_A$ values (derived from simulations):
for galaxies chosen by stellar mass \citep{Barreira:2020kvh},
\begin{equation}
p_{g} \approx 0.55\;,\label{e2.7}
\end{equation}
and for HI intensity mapping \citep{Barreira:2021dpt}
\begin{equation}
p_{\rm HI} \approx 1.25\;.\label{e2.8}
\end{equation}
 {In fact, the galaxy samples we use are not chosen by stellar mass but we are not aware of simulations to estimate $p_g$ values for these samples. Consequently we consider two further values:
\begin{equation}
p_{g} \approx 1\quad \text{and}\quad 1.5\;,\label{e2.9}
\end{equation}
and we find the effect on $\sfnl$ of these changes.}


\section{Surveys}\label{Surveys}

We consider two pairs of nominal spectroscopic surveys, each pair with the same redshift range and with sky area $\Omega_{\rm sky} =10,000~ {\rm deg}^2$:
\begin{itemize}
\item  galaxy survey `H$\alpha$', and HI intensity mapping survey `H', $1\leq z\leq 2$ (similar to $Euclid$ \cite{Euclid:2019clj} and HIRAX \cite{Newburgh:2016mwi,Crichton:2021hlc} respectively);
\item galaxy survey `LBG' and  HI intensity mapping survey `P', $2\leq z\leq 5$  (similar to   MegaMapper \citep{Schlegel:2022vrv} and PUMA \cite{PUMA:2019jwd} respectively).
\end{itemize}

\begin{figure}[! ht]
\centering
\includegraphics[width=7.6cm]{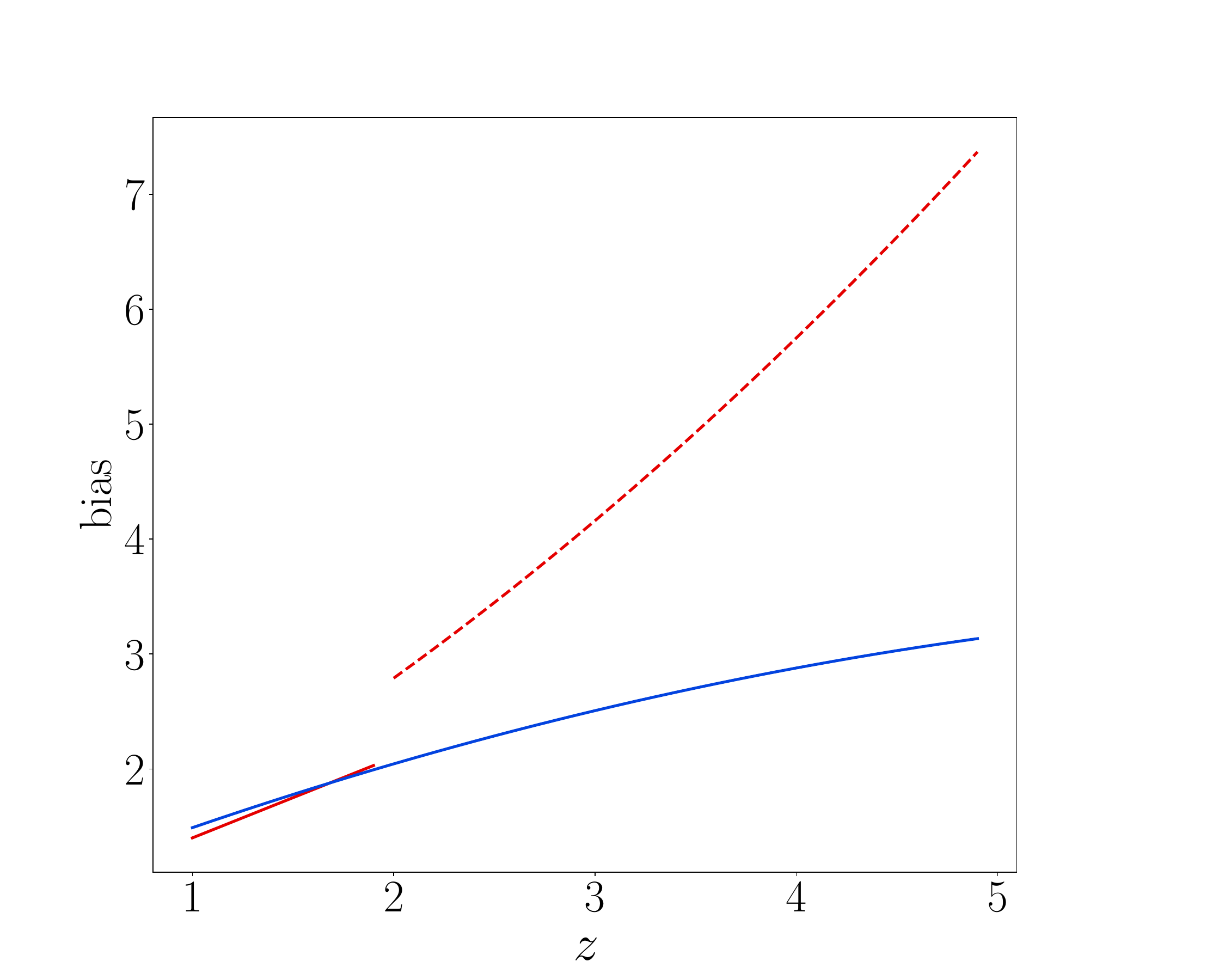} 
\includegraphics[width=7.6cm]{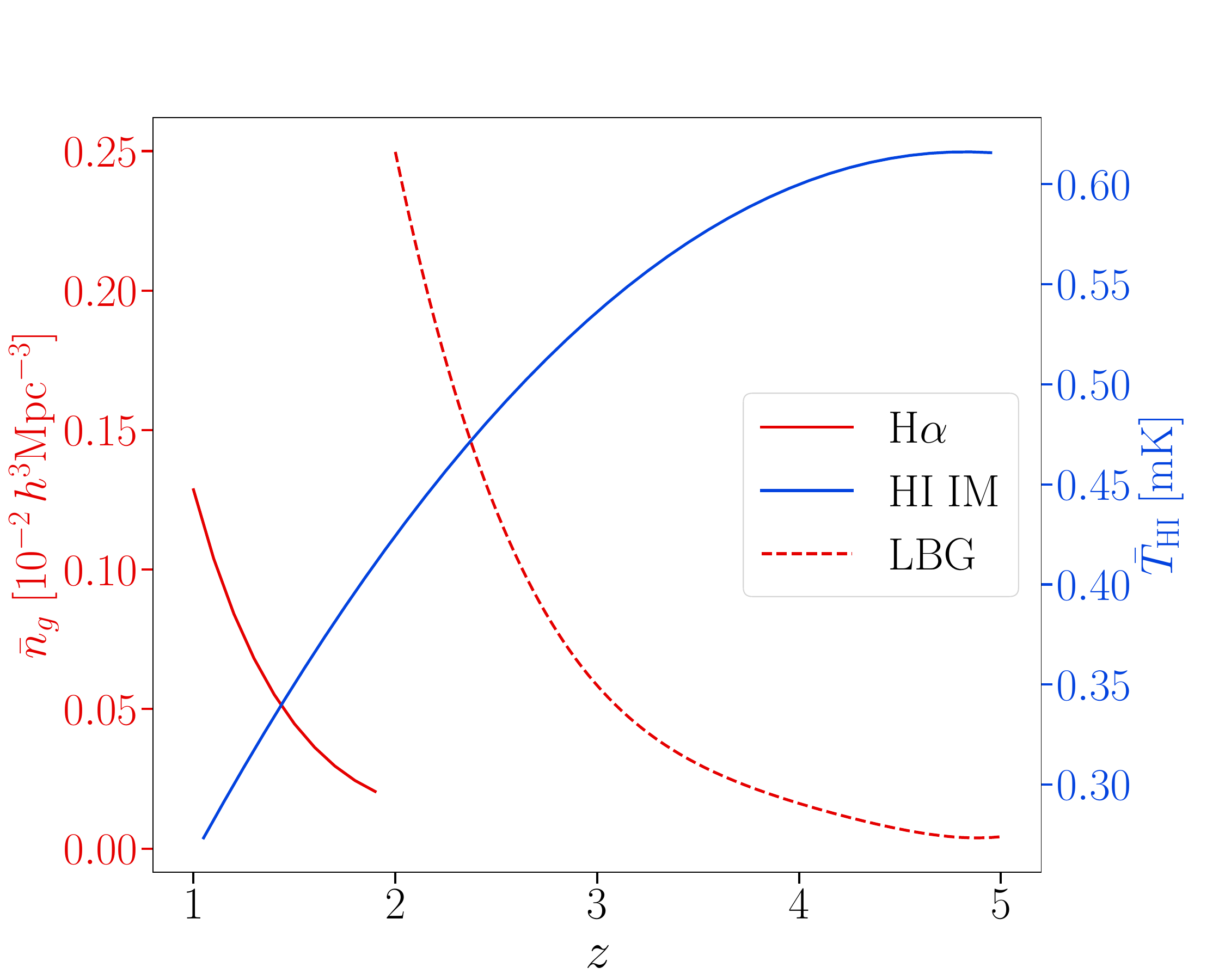} 
\caption{For the galaxy and HI IM surveys: clustering biases ({\em left panel}); 
comoving number densities 
{({\em right panel}, red, left-hand $y$-axis)}
and HI IM temperature ({\em right panel}, blue, right-hand $y$-axis).
} \label{fig1}
\end{figure}

\noindent The Gaussian clustering biases are as follows:
\begin{align}
b_{{\rm H}\alpha}(z) &= b_{{\rm H}\alpha 0}(1+z) & \text{fiducial}~b_{{\rm H}\alpha 0}=0.7~~~~\text{\citep{Merson:2019vfr}},
\label{e3.1}
\\ 
b_{\rm LBG}(z) &= b_{\rm LBG 0}\big(1 +1.155\,z + 0.155\,z^{2}\big) & \text{fiducial}~ b_{\rm LBG 0}=0.710~\text{\citep{Sailer:2021yzm}},
\label{e3.2}\\
b_{\rm H}(z) &=b_{\rm P}(z) =b_{\rm HI}(z)= b_{\rm HI 0}\big(1 + 0.823 z - 0.0546 z^{2}\big) & \text{fiducial}~ b_{\rm HI 0}=0.842 ~\text{\citep{Villaescusa-Navarro:2018vsg}}.\label{e3.3}
\end{align}
The background comoving number densities for   H$\alpha$  \cite{Maartens:2021dqy} and LBG \citep{Ferraro:2019uce} are taken as
\begin{align}
\bar n_{{\rm H}\alpha}(z) &= \big(3.63\, z^{-0.910}{\rm e}^{0.402\,z} - 4.14\big) \times 10^{-3}h^{3}~\mathrm{Mpc^{-3}}\,, \label{e3.4}
\\
\bar n_{\rm LBG}(z) &= \big(2.43 - 2.21\,z  +0.773\,z^{2} - 0.122\,z^{3}+ 0.00733\,z^{4}\big) \times 10^{-2}h^{3}\,\mathrm{Mpc^{-3}}\,. \label{e3.5}
\end{align}
For HI IM, the background brightness temperature is modelled as \citep{Santos:2017qgq}:
\begin{equation}
\bar{T}_{\rm HI}(z) = 0.0559 +0.2324\,z -0.0241\, z^{2} ~~ \mathrm{mK}\,. \label{e3.6}
\end{equation} 
Equations \eqref{e3.1}--\eqref{e3.6} are shown in  \autoref{fig1}. 

\subsection{Noise}\label{Noise}

For galaxy surveys, $g={\rm H}\alpha$ or LBG, the shot noise is assumed to be Poissonian, so that the total observed auto-power is
\begin{equation}
\tilde{P}_{g}(z,k,\mu) = P_{g}(z,k,\mu) +  \frac{1}{\bar n_{ g}(z)}\;. \label{e3.7}
\end{equation}
In HI IM surveys, HI\,=\,H or P, the dominant noise on linear scales is the thermal (or instrumental) noise \cite{Castorina:2016bfm, Villaescusa-Navarro:2018vsg}, so that shot noise can safely be neglected and the observed HI IM power spectrum is  
\begin{equation}
\tilde{P}_{\rm HI}(z,k,\mu) = P_{\rm HI}(z,k,\mu) + P_{\rm HI}^{\mathrm{therm}}(z,k,\mu)\;,\label{e3.12}
\end{equation}
where $P_{\rm HI}^{\mathrm{therm}}$ depends on the model of the sky temperature in the radio band, the survey specifications and the survey mode (single-dish or interferometer). For the interferometer (IF) mode that we consider, the thermal noise power spectrum is \cite{Bull:2014rha,Alonso:2017dgh,Ansari:2018ury,Jolicoeur:2020eup}:
\begin{equation}
P_{\rm HI}^{\mathrm{therm}}(z,k_\perp) = \frac{\Omega_{\mathrm{sky}}}{2\nu_{21} t_{\mathrm{tot}}}\,\frac{(1+z) r(z)^{2} }{\cH(z) }\,\left[\frac{T_{\mathrm{sys}}(z)}{\bar{T}_{\rm HI}(z)}\right]^2\, \left[\frac{\lambda_{21}(1+z)}{A_{\mathrm{e}}}\right]^2\frac{1}{n^{\mathrm{phys}}_{\mathrm{b}}(z,k_{\perp})}
{\frac{1}{\theta_{\rm fov}(z)^2}}
\;,  
\label{e3.8}
\end{equation}
where $k_\perp=k\sqrt{1-\mu^2}$ is  the transverse mode, $\nu_{21}=1420\,\mathrm{MHz}$ is the rest-frame frequency of the $\lambda_{21}=21\,$cm emission,  $t_\mathrm{tot}$ is the total observing time, $r$ is the radial comoving distance, $\cH$ is the conformal Hubble rate, 
 {and the system temperature is modelled as \citep{Ansari:2018ury}:
\begin{equation}
T_{\mathrm{sys}}(z) = T_{\rm d}(z)+T_{\rm sky}(z) =T_{\rm d} + 2.7 + 25\bigg[\frac{400\,\mathrm{MHz}}{\nu_{21}} (1+z)\bigg]^{2.75} ~ \mathrm{K}, \label{e3.11}
\end{equation} 
where $T_{\rm d}$ is the dish receiver temperature, given in \autoref{tab2}.}
The effective beam area is
\begin{equation}
A_{\mathrm{e}} = 0.7\, \frac{\pi}{4}D_{\mathrm{d}}^{2}\;,\label{e3.9} 
\end{equation}
with dish diameter $D_{\mathrm{d}}$ and 
the field of view is
\begin{equation}
\theta_{\rm fov}(z) = 1.22\,\frac{\lambda_{21}(1+z)}{D_{\rm d}}\;. \label{e3.14}
\end{equation}

The physical baseline density of the array distribution is modelled following
\cite{Ansari:2018ury}.
For a given baseline of length $L$, we use the fitting formula for close-packed arrays given in \cite{Ansari:2018ury} (see also \cite{Karagiannis:2019jjx,Jolicoeur:2020eup}): 
\begin{equation}
n^{\mathrm{phys}}_{\mathrm{b}}(z,k_{\perp}) = \left(\frac{N_{\mathrm{s}}}{D_{\mathrm{d}}}\right)^{2}\frac{a+b(L/L_{\mathrm{s}})}{1+c(L/L_{\mathrm{s}})^{d}}\exp{\big[-(L/L_{\mathrm{s}})^{e}\big]} ~~ \text{with} ~~ L = \frac{k_{\perp}r(z)}{2\pi}\lambda_{21}(1+z)\,,\label{e3.10}
\end{equation} 
where $L_{\mathrm{s}}=N_{\mathrm{s}}D_{\mathrm{d}}$. 
H  {(HIRAX-like)} is a square close-packed array,  while P  {(PUMA-like)} is a hexagonal close-packed array with 50\% fill factor.
This means that half the dish sites are randomly empty, so the array is equivalent in size to that of twice the number of  {dish elements $N_{\rm d}$} but with a quarter of the baseline density \cite{Karagiannis:2019jjx}.
Then it follows that $N_{\rm s}^2 =N_{\rm d}$ for H and $N_{\rm s}^2 =2N_{\rm d}$ for P \citep{Ansari:2018ury, Castorina:2020zhz}.
The values of the fitting parameters  {$a,b,c,e$ in \eqref{e3.10} are given in \autoref{tab1}, from \cite{Ansari:2018ury} (their Appendix D, last equation).} 

\begin{figure}[! h]
\centering
\includegraphics[width=7.6cm]{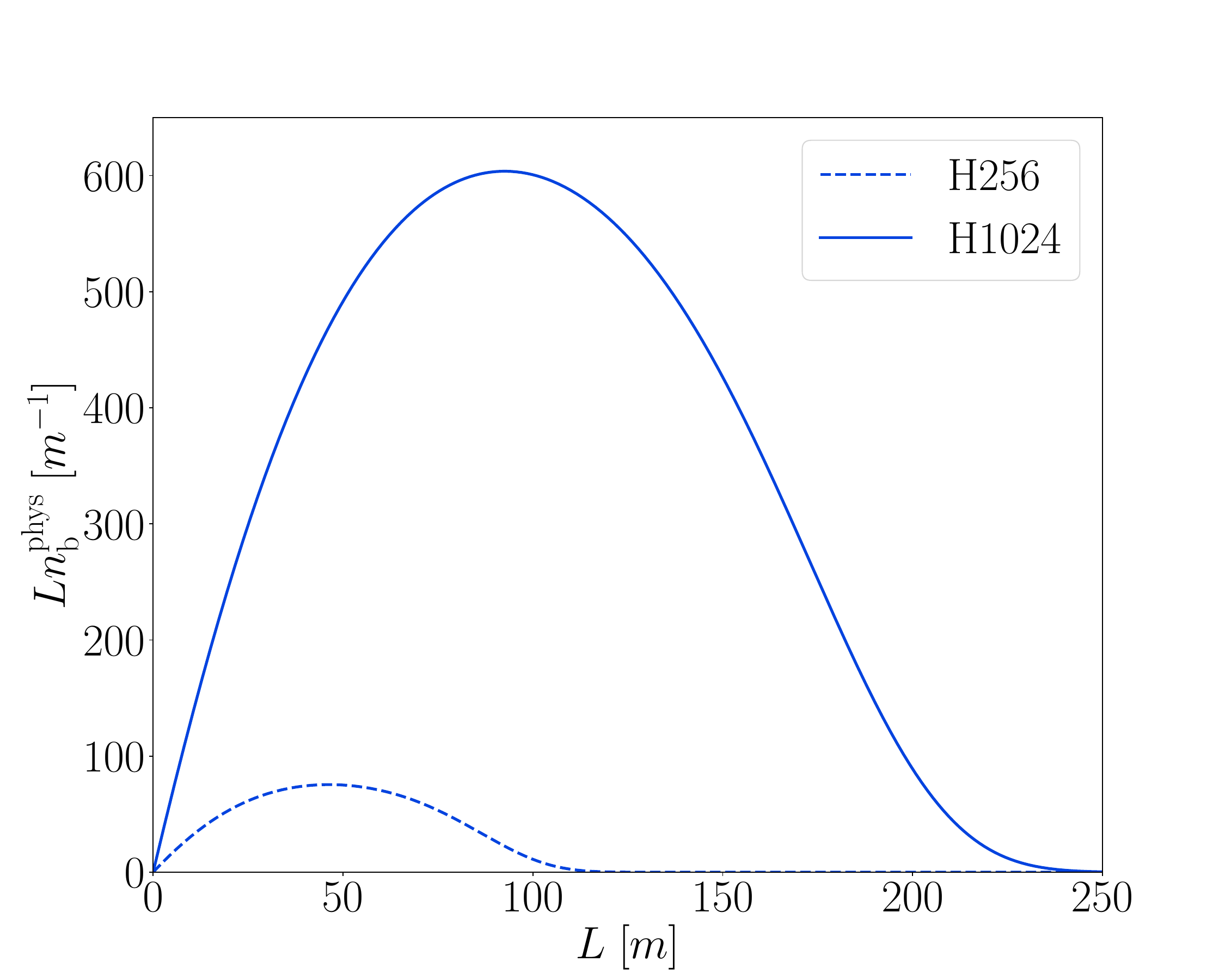} 
\includegraphics[width=7.6cm]{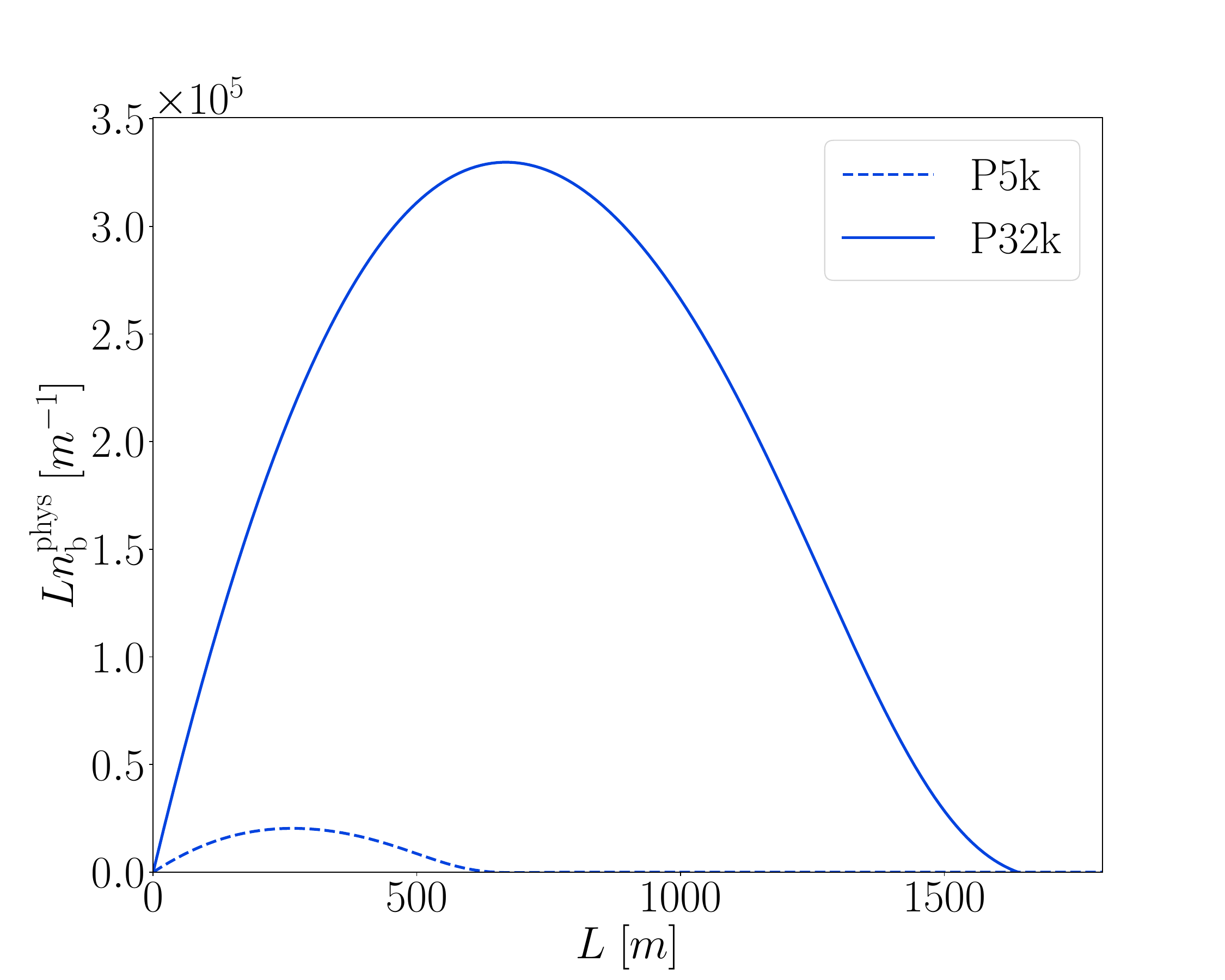} 
\caption{ Physical baseline density for HI IM surveys:  {HIRAX-like  (\emph{left}) and PUMA-like  (\emph{right})}.
} \label{fig2}
\end{figure}

The baseline densities $L n^{\mathrm{phys}}_{\mathrm{b}}$ for the H and P surveys are shown in \autoref{fig2}.
Note that the thermal noise \eqref{e3.8}, which is $\propto 1/n_{\rm b}^{\rm phys}$, tends to infinity at the maximum baseline $D_{\rm max}$.

\begin{table}[!ht] 
\centering 
\caption{Specifications of HI IM surveys.  {H is HIRAX-like, with 256 dishes in phase 1 and 1024 in phase 2; P is PUMA-like, with 5k dishes in phase 1 and 32k in phase 2. $N_{\rm d}$ is the number of dishes, $D_{\rm d}$ is the dish diameter, $T_{\rm d}$ is the dish receiver temperature, $D_{\rm max}$ is the maximum baseline, $t_{\rm tot}$ is the observing time. The parameters $N_{\rm s}$ and $a,\cdots, e$ are given in \eqref{e3.10}}.\\}
 \label{tab1} 
\vspace*{0.2cm}
\begin{tabular}{|l|c|c|c|c|c|c|} 
\hline 
Survey~~&~~$N_{\mathrm{d}}$~~& ~~$N_{\mathrm{s}}$~~&~~$t_{\rm tot}$~~~&~~$T_{\mathrm{d}}$~~& ~~$D_{\mathrm{d}}$~~& ~~$D_{\mathrm{max}}$~~ \\ 
 & & &~~[$10^{3}\rm{hr}$]&~~[K]~~&~~[m]~~&~~[m]~~ \\ \hline \hline
H\,256,\,1024 & 256,\,1024 & 16,\, 32 & 12 & 50 & 6 & 141,\, 282~~ \\
P\,5k,\,32k & 5k,\,32k &  100,\,253 & 20 & 93 & 6 & 648,\,1640 \\ \hline \\ \hline 
Fitting &~~$a$~~ &~~$b$~~ &~~$c$~~&~~$d$~~&~~$e$~~& \\
parameters & & & & & & \\
\hline \hline
H\,256,\,1024 & 0.4847 & -0.3300 & 1.3157 & 1.5974 & 6.8390~~& \\ 
P\,5k,\,32k & 0.5698 & -0.5274 & 0.8358 & 1.6635 & 7.3177~~&  \\ \hline 
\end{tabular}
\end{table}

\begin{figure}[!h]
\centering
\includegraphics[width=8.6cm]{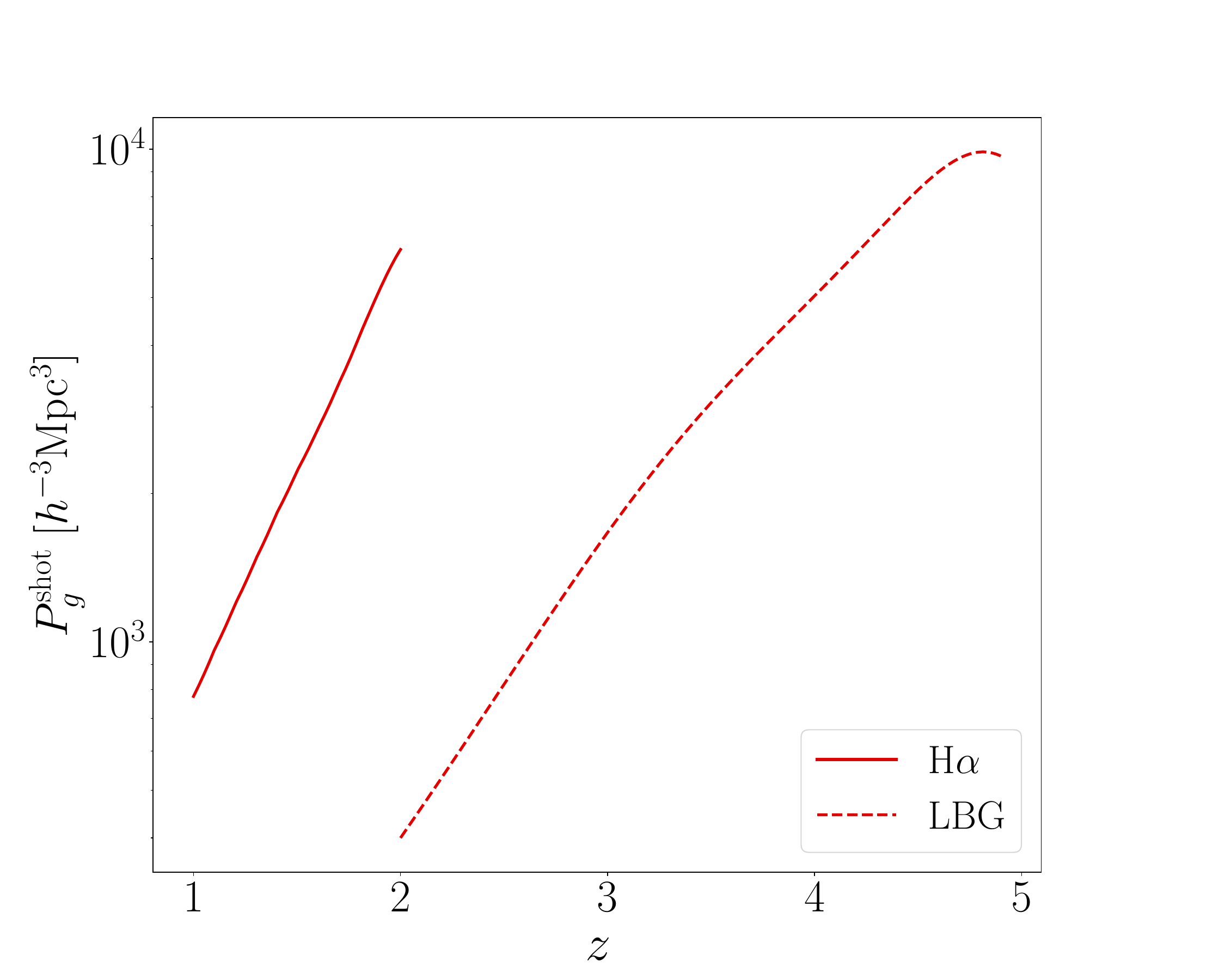} 
\caption{ {Redshift evolution of galaxy shot noise.} 
} \label{fig3}
\end{figure} 
\begin{figure}[! h]
\centering
\includegraphics[width=7.6cm]{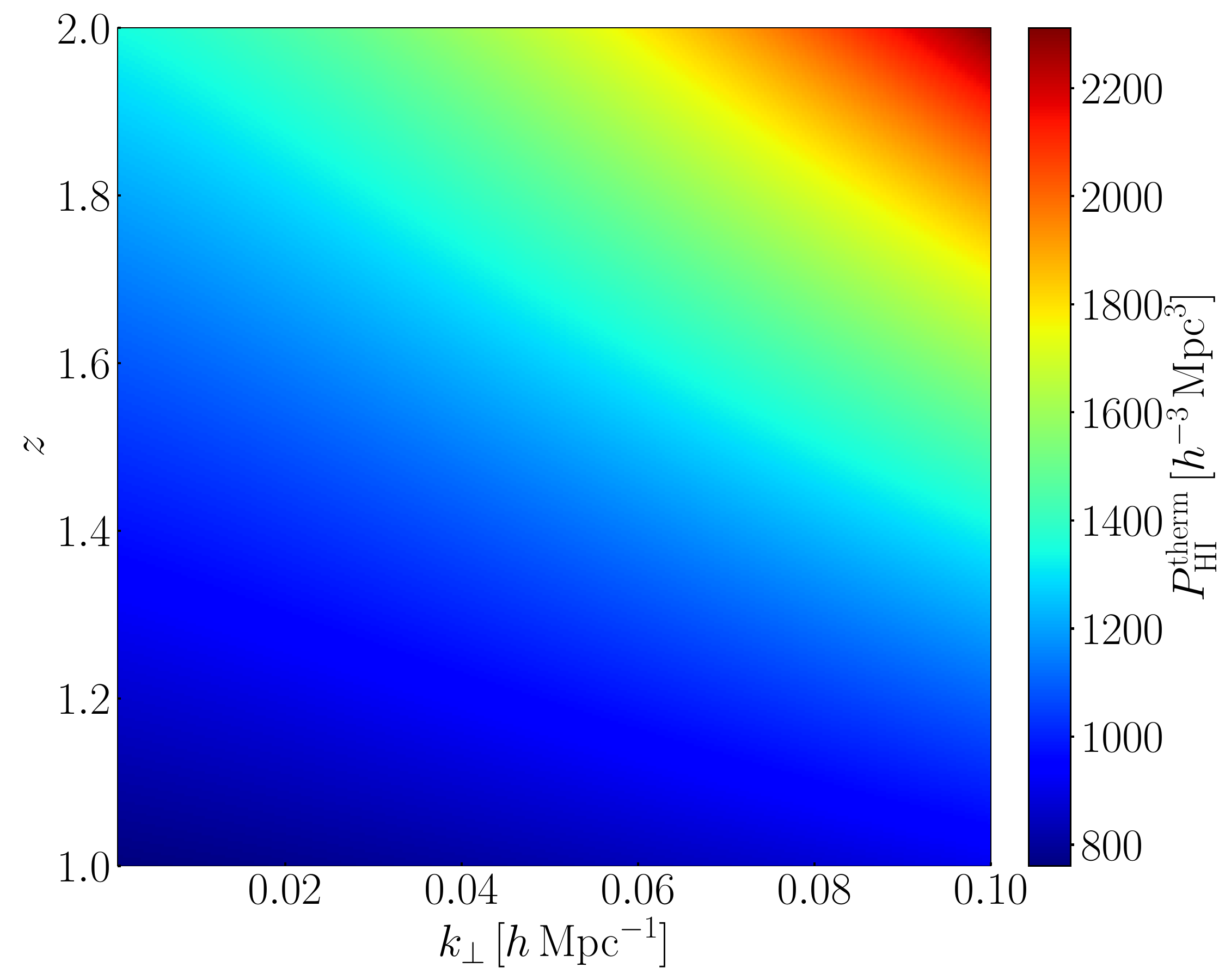} 
\includegraphics[width=7.6cm]{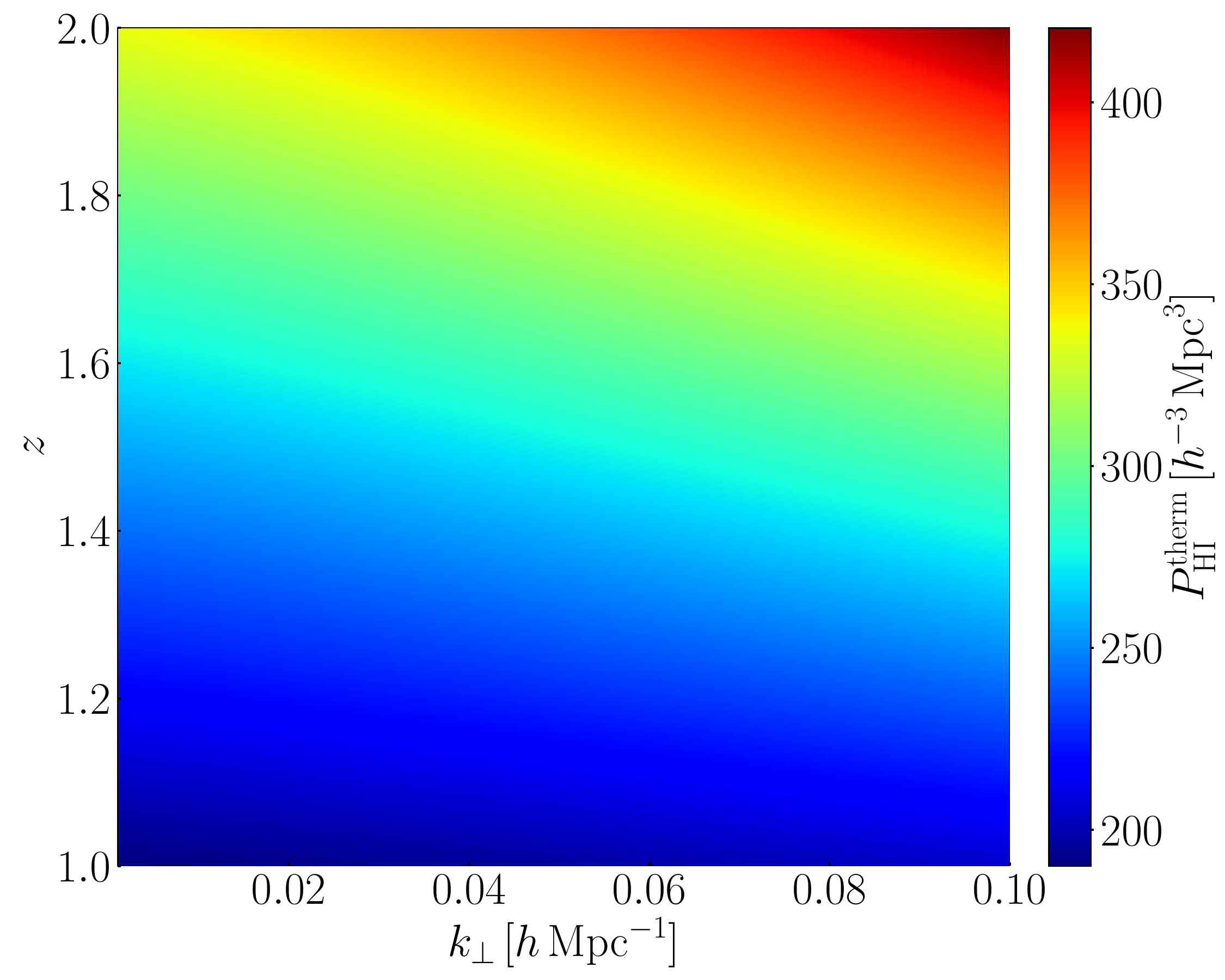} \\
\includegraphics[width=7.6cm]{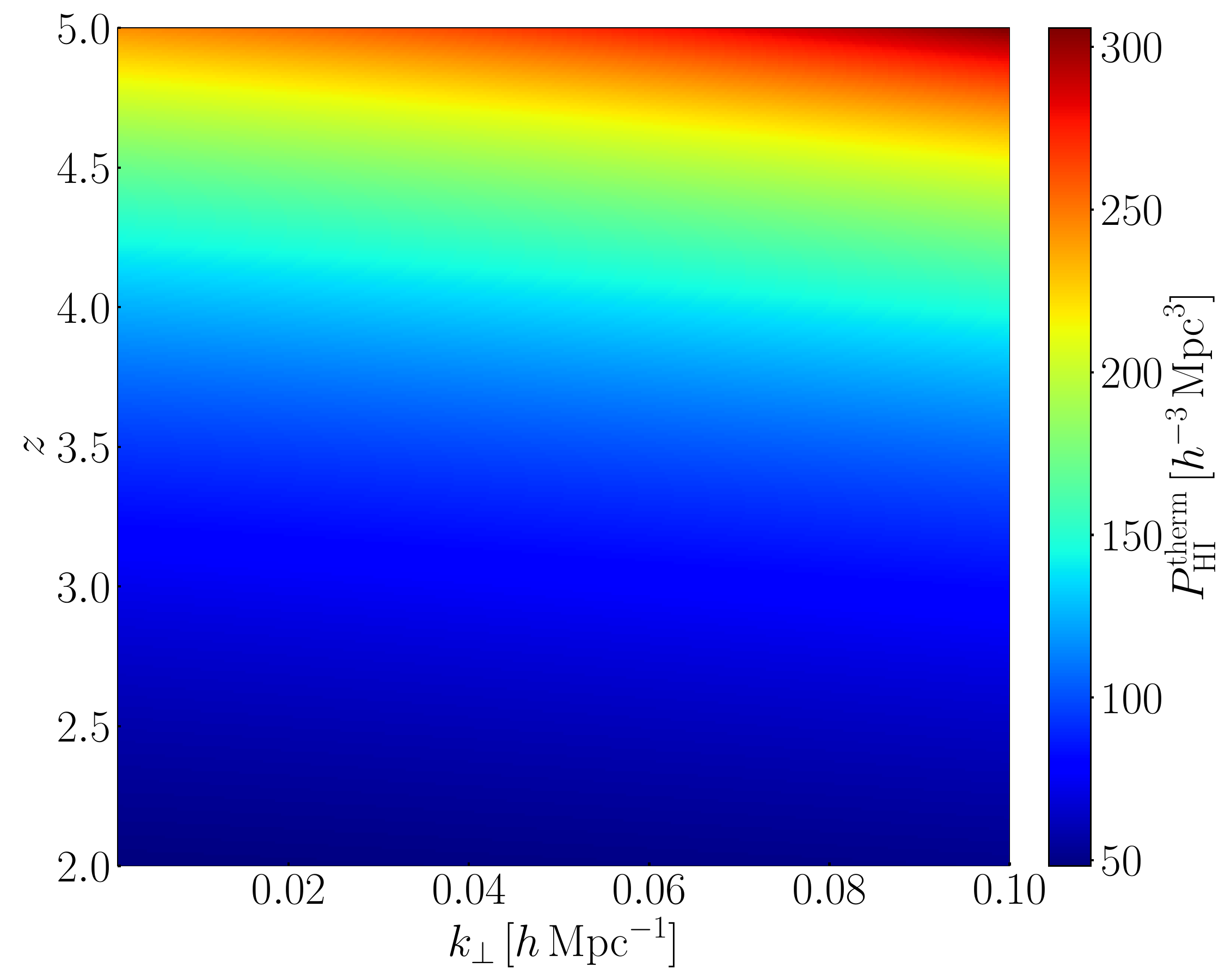} 
\includegraphics[width=7.6cm]{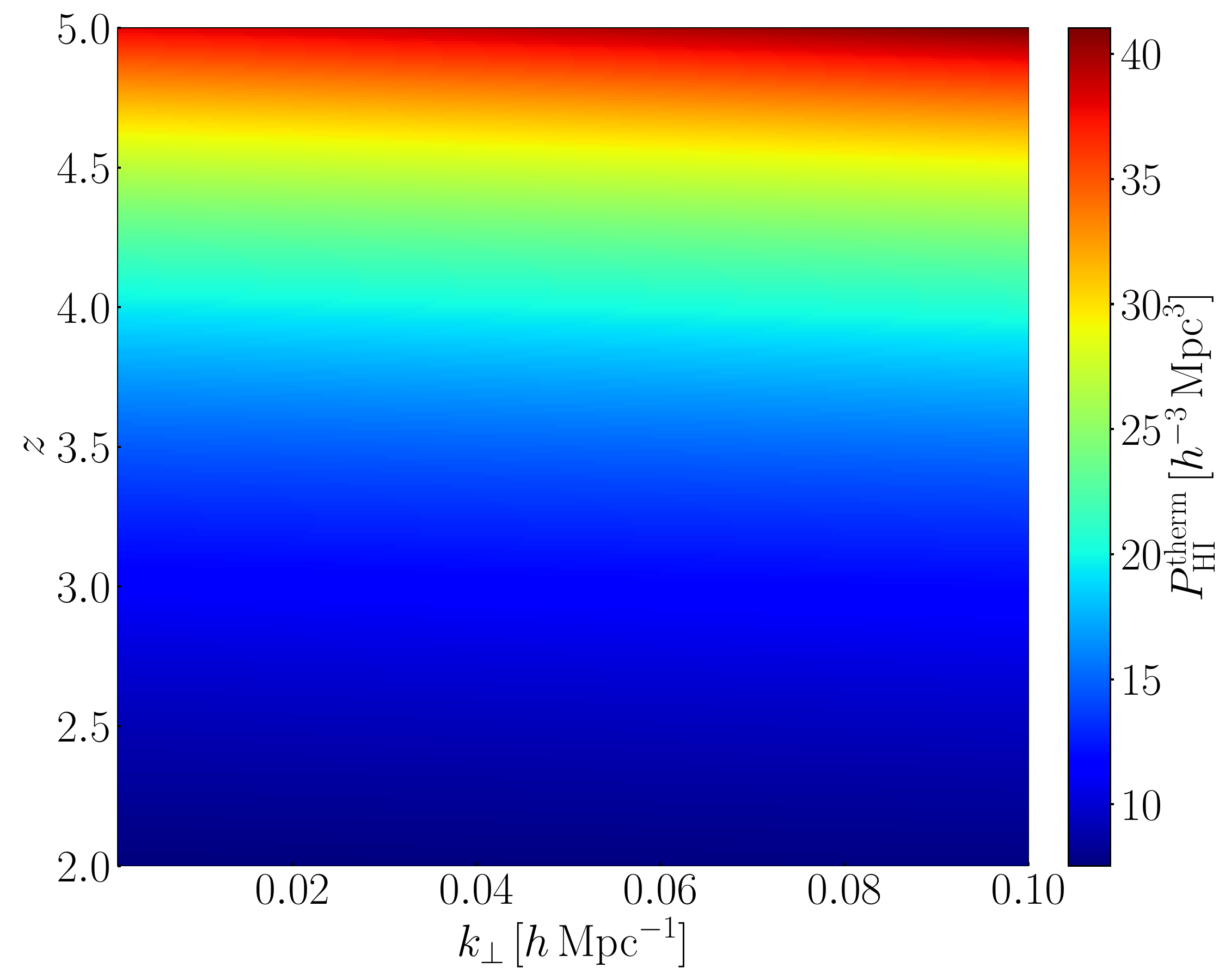}
\caption{2D colour plot of the scale- and redshift-dependence of thermal noise for the HI IM surveys:  {HIRAX-like (256 dishes) (\emph{top left}); HIRAX-like (1024 dishes) (\emph{top right}); PUMA-like (5k dishes) (\emph{bottom left}); PUMA-like (32k dishes) (\emph{bottom right}).}
} \label{fig2_11}
\end{figure}

The cross-noise power spectrum between HI IM and galaxy surveys depends on the extent of overlapping halos that host the HI emitters and the galaxies. Typically it is neglected, as motivated in \cite{Viljoen:2020efi,Casas:2022vik}. Then the observed cross-power spectra are 
\begin{equation}
\tilde P_{g{\rm HI}}=P_{g{\rm HI}}\,. \label{e3.13}
\end{equation}
The noise auto-power for  {galaxy} surveys is shown in \autoref{fig3}.
  {The HI IM thermal noise  power spectra are displayed in \autoref{fig2_11}.}

\subsection{HI IM  foreground avoidance}\label{HI IM  foreground avoidance}

HI IM surveys are contaminated by foregrounds that are much larger than the HI signal. The cleaning of foregrounds from the HI signal relies on the fact that the foregrounds are nearly spectrally smooth (see e.g. \cite{Cunnington:2020wdu,Spinelli:2021emp} and references there in for recent advances). However,  the cleaning is not perfect, and in some regions of Fourier space the signal loss is large.  For simplified Fisher forecasts, we assume that cleaning is successful, except in the known  regions of Fourier space where cleaning is  expected to be least effective. Then we use  foreground filters that avoid these regions. There are two regions of foreground avoidance for IF-mode surveys. 

\begin{enumerate}
\item 
We filter out long-wavelength radial modes  with $|k_{\parallel}|\lesssim k_{\rm fg}$, where $k_{\rm fg}$ is a suitable minimum radial mode for future surveys.
We follow 
\cite{Cunnington:2021czb} and use the radial damping factor    
\begin{equation}
{\cal D}_{\rm fg}(k,\mu) = 1 - \exp\bigg[-\bigg(\frac{k_{\parallel} } {k_{\rm fg}}\bigg)^{\!\!{2}\,}\bigg]\;.  \label{e3.15} 
\end{equation}
 {In order to find the effect on $\fnl$ constraints of changing $k_{\rm fg}$, we consider a fiducial value that follows \cite{Cunnington:2021czb} and a less optimistic value:
\begin{align}\label{kfgflo}
   k_{\rm fg} = 0.005 ~ h{\rm Mpc}^{-1}~~\text{(fiducial)}\,,\quad
  k_{\rm fg} = 0.01 ~ h{\rm Mpc}^{-1}~~\text{(less optimistic)}\,.
\end{align}}

\item
A physical baseline in the IF array probes different angular scales $k_\perp$ at different frequencies (and hence different $z$). As a consequence,  monochromatic emission from a foreground point source can contaminate the signal in a wedge-shaped region of Fourier space \cite{Pober:2013jna,Pober:2014lva}. This contamination occurs within $N_{\rm w}$ primary beams of a pointing, where $N_{\rm w}=0$ is the ideal case of no contamination, which occurs in principle if  calibration of the interferometer is near-perfect \cite{Ghosh:2017woo}. Currently this precision in calibration and stability is not achievable. In order to avoid the wedge region we therefore impose the filter \cite{Pober:2013jna,Pober:2014lva,Obuljen:2017jiy,Alonso:2017dgh,Ansari:2018ury}
\begin{equation}
\left|k_{\parallel}\right| > A_{\mathrm{w}}(z) \, k_{\perp} \quad \text{where} \quad A_{\mathrm{w}}(z) = r(z) \mathcal{H}(z) \sin\big[0.61 N_{\mathrm{w}} \,\theta_{\mathrm{fov}}(z)\big]\;. \label{e3.16}
\end{equation}
 {We investigate the effect on $\fnl$ constraints of changing $N_{\mathrm{w}}$ by using a fiducial value  and a less optimistic value,  following \cite{Karagiannis:2019jjx}:
\begin{align}\label{nwflo}
   N_{\mathrm{w}} = 1~~\text{(fiducial)}\,,\quad
  N_{\mathrm{w}}  = 3~~\text{(less optimistic)}\,.
\end{align}
The fiducial value corresponds to foreground contamination from sources within 1 primary beam of the pointing, while the less optimistic value means that contamination extends to 3 primary beams.} 
\end{enumerate}

\begin{figure}[! ht]
\centering 
\includegraphics[width=7.6cm]{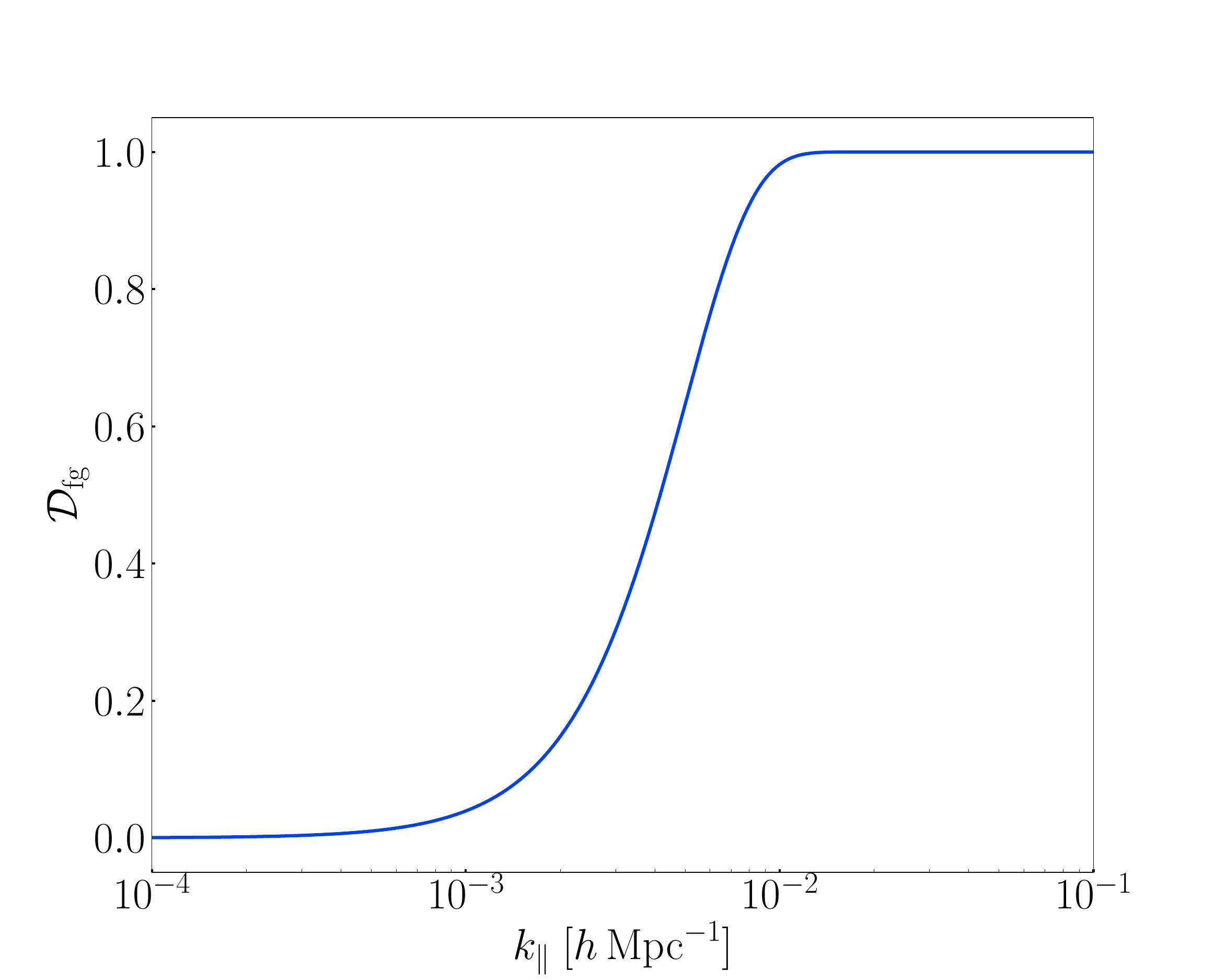} 
\includegraphics[width=7.6cm]{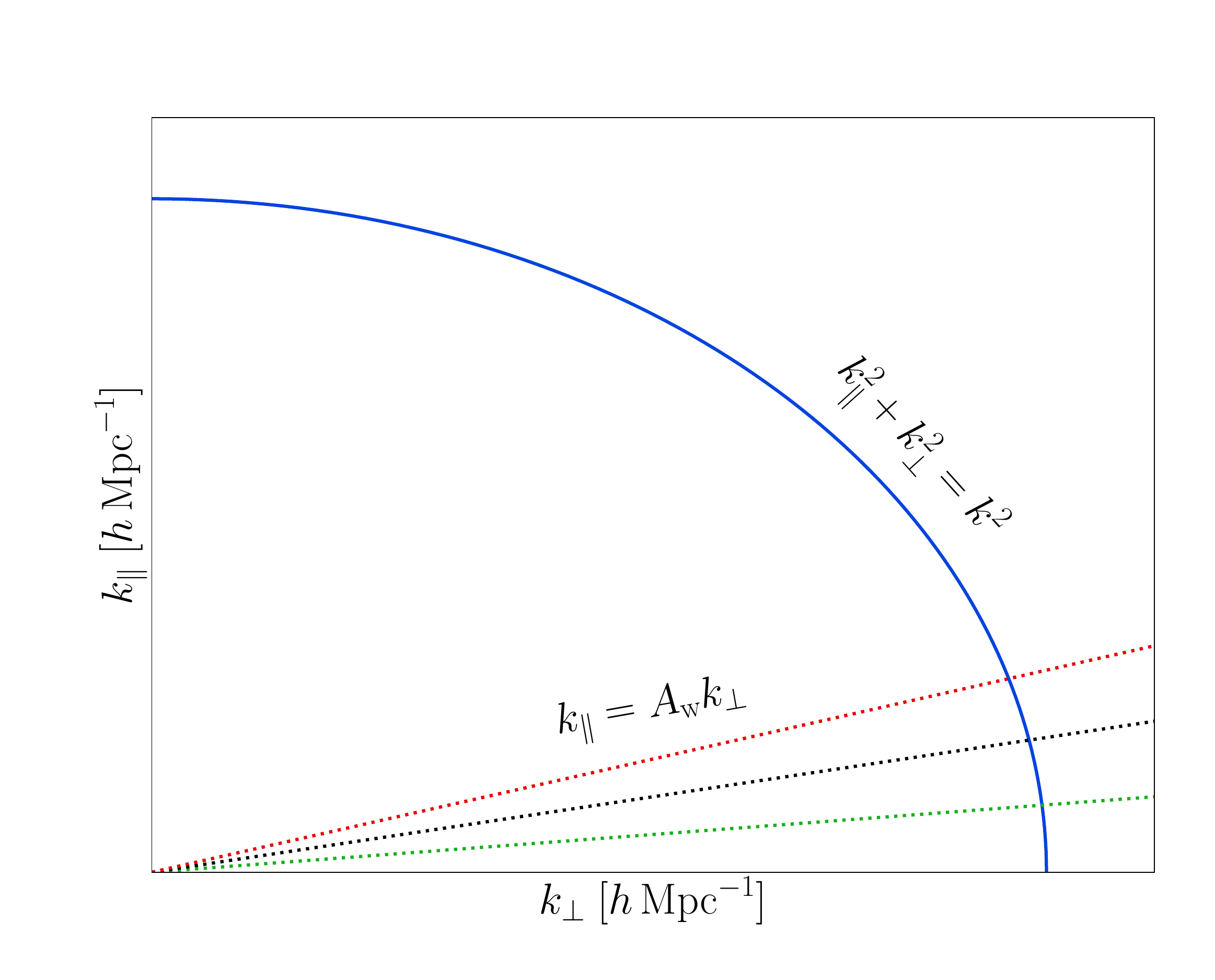} 
\caption{Radial foreground damping function (\emph{left}). Schematic to illustrate the foreground wedge; the wedge angle increases with redshift (\emph{right}).
}
\label{fig4}
\end{figure}

\autoref{fig4} illustrates ${\cal D}_{\rm fg}$ and $A_{\mathrm{w}}$.
Foreground filters lead to the following modifications of the HI IM  {temperature contrasts: 
\begin{eqnarray}
\delta_{\rm HI}(z,{k}, \mu) \rightarrow  {\cal D}_{\rm fg}(k,\mu)
\,\Theta\big(\left|k_{\parallel}\right| - A_{\mathrm{w}}(z) k_{\perp} \big)
\,\delta_{\rm HI}(z,{k}, \mu)\,, \label{e3.18} 
\end{eqnarray}
where $\Theta$ is the Heaviside step function. Note that
$P_{\rm HI}\to {\cal D}_{\rm fg}^2\Theta P_{\rm HI}$ and 
$P_{g\rm HI}\to {\cal D}_{\rm fg}\Theta P_{g\rm HI}$.}


\subsection{Maximum and minimum scales}\label{Maximum and minimum scales}

Since the local PNG signal in power spectra is on ultra-large scales, we only require linear perturbation theory. This means that we can exclude scales where linear perturbations break down, i.e.  $k<k_{\mathrm{max}}$, where \citep{Smith:2002dz,Maartens:2019yhx,Fonseca:2019qek,Modi:2019hnu}
\begin{equation}
k_{\mathrm{max}}(z) = k_{\mathrm{max},0}\big(1+z\big)^{2/(2+n_{s})}\;. \label{e3.21}
\end{equation} 
Then we choose a very conservative value of $k_{\mathrm{max},0} = 0.08\;h\,\mathrm{Mpc}^{-1}$. 

\begin{figure}[! ht]
\centering 
\includegraphics[width=7.6cm]{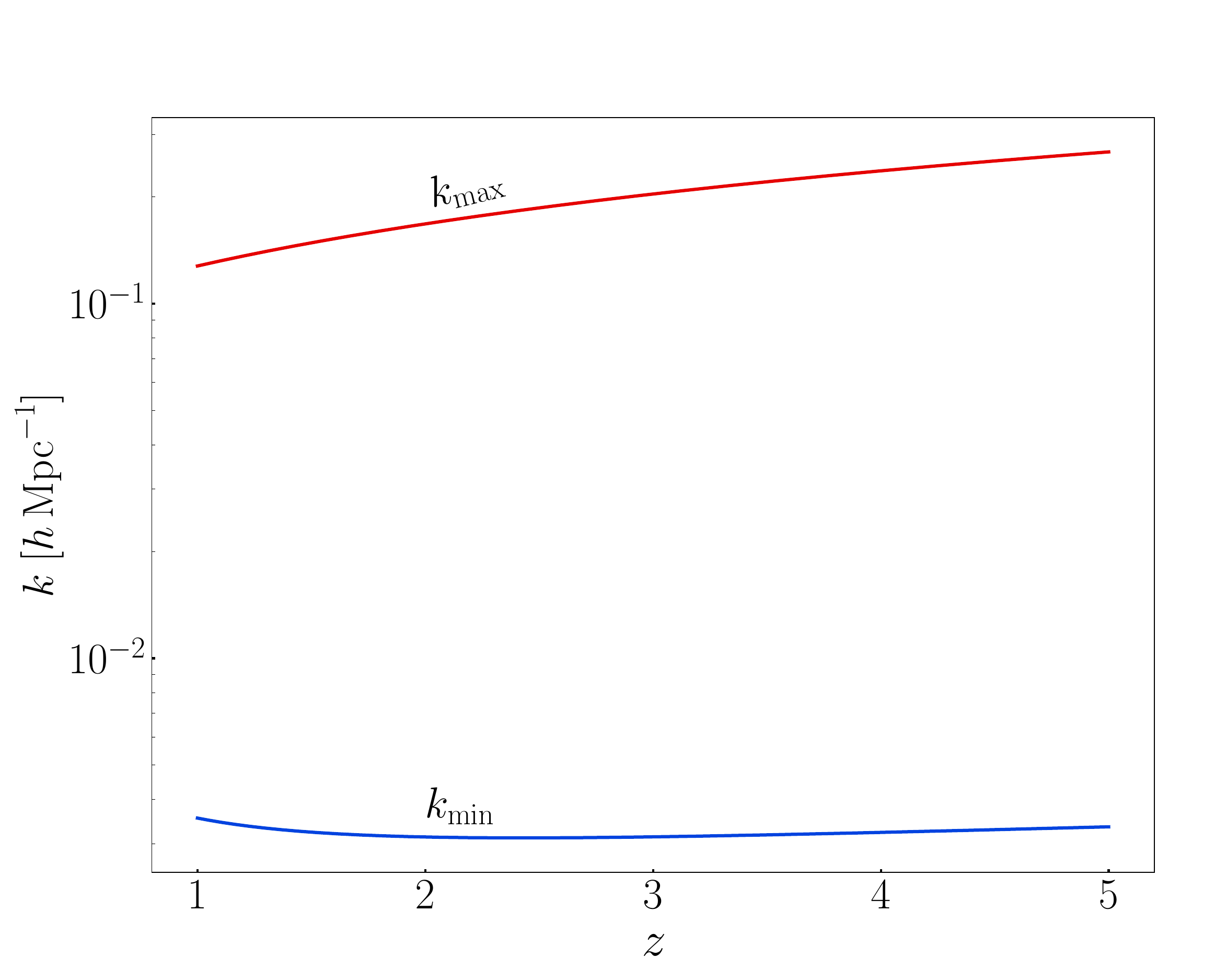} 
\includegraphics[width=7.6cm]{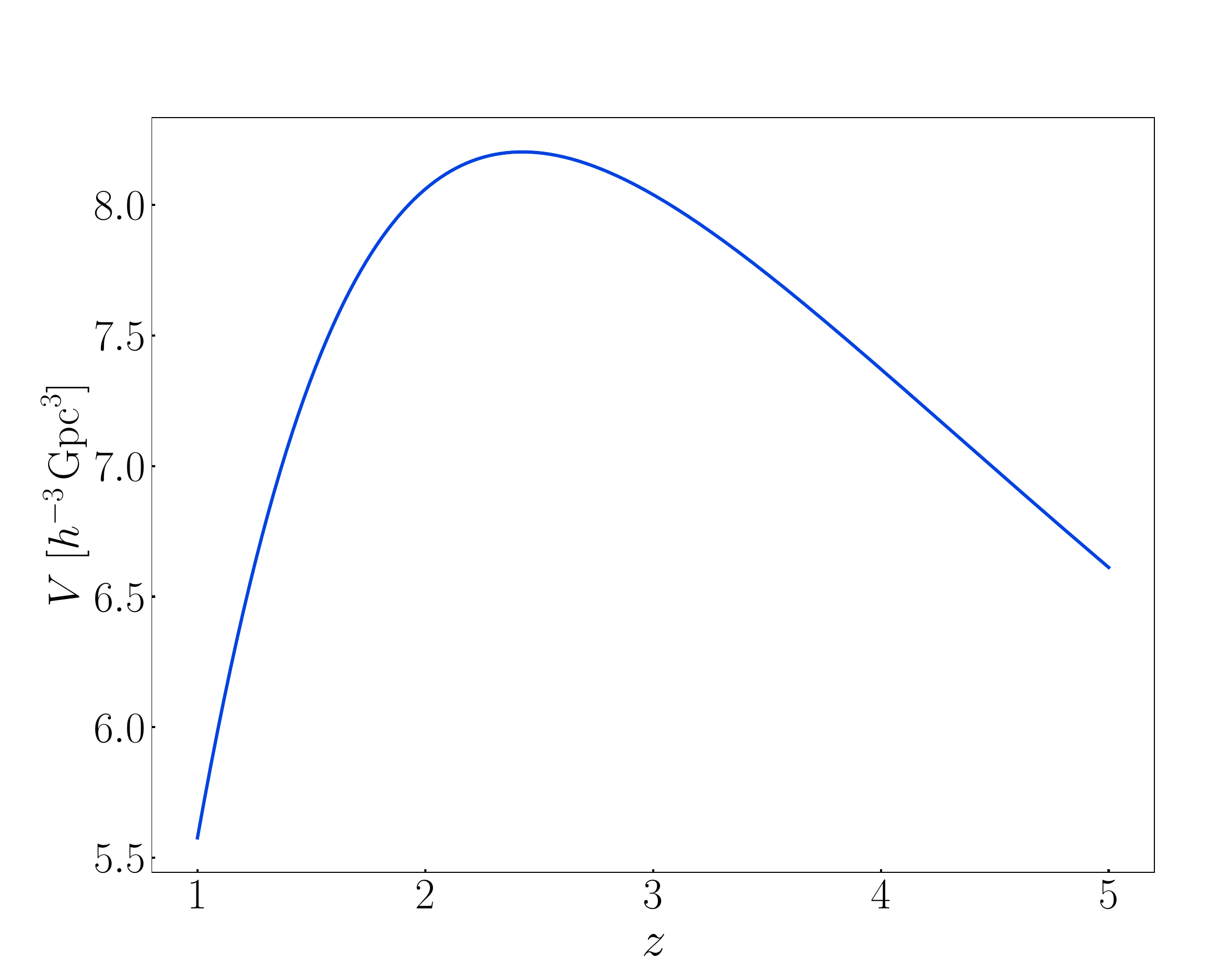}  
\caption{Minimum and maximum wavenumbers ({\em left}) and comoving volume  ({\em right}) over the redshift range of the surveys.
} \label{fig5}
\end{figure}

The minimum wavenumber $k_{\mathrm{min}}$ in each redshift bin is the fundamental mode $k_{\mathrm{f}}$, given by the comoving volume of the bin of width  {$\Delta z=0.1$} centred at $z$:   
\begin{equation}
k_{\mathrm{min}}(z)=k_{\mathrm{f}}(z) = \frac{2\pi}{V(z)^{1/3}} \quad \text{with} \quad V(z) = \frac{\Omega_{\mathrm{sky}}}{3}\Big[r\big(z+\Delta z/2\big)^{3} - r\big(z-\Delta z/2\big)^{3}\Big]\;.\label{e3.23}
\end{equation}
The minimum and maximum scales are shown in \autoref{fig5}.


\section{Fisher forecast}\label{Fisher forecast}

For a given redshift bin, the multi-tracer Fisher matrix for the combination of two dark matter tracers  $g$ and HI is 
\begin{equation}
F_{\alpha \beta}^{\bm{P}} = \sum_{\mu=-1}^{+1}\,\sum_{k=k_{\mathrm{min}}}^{k_{\mathrm{max}}}\,\partial_{\alpha}\,\bm{P} \cdot \mathrm{Cov}(\bm{P},\bm P)^{-1} \cdot \partial_{\beta}\,\bm{P}^{\mathrm{T}}\;, \label{e4.1}
\end{equation}
where $\partial_{\alpha} = \partial\,/\,\partial \vartheta_{\alpha}$  {and $\vartheta_{\alpha}$ are the parameters to be constrained}. Here $\bm P$ is the data vector of the power spectra, 
\begin{equation}
\bm{P} = \big( P_{g}\,,\,  P_{g{\rm HI}}\,,\,  P_{\rm HI} \big)\;.\label{e4.2}
\end{equation}
Note that the noise is  excluded since it is independent of the cosmological and nuisance parameters. The noise enters the covariance \cite{White:2008jy,Zhao:2020tis,Barreira:2020ekm,Karagiannis:2023lsj}: 
\begin{equation} 
\mathrm{Cov}(\bm{P},\bm P) =  \frac{k_{\rm f}^3}{4\pi k^{2} \Delta k}\,\frac{2}{\Delta \mu}\,
\begin{pmatrix}
\tilde P_{g}^2 & & \tilde P_{g}\tilde P_{g{\rm HI}} & & \tilde P_{g{\rm HI}}^2 \\ 
& & & & \\
\tilde P_{g}\tilde P_{g{\rm HI}} & & \frac{1}{2}\big(\tilde P_{g}\tilde P_{\rm HI}+ \tilde P_{g{\rm HI}}^2 \big) & & \tilde P_{\rm HI}\tilde P_{g{\rm HI}} \\
& & & & \\
\tilde P_{g{\rm HI}}^2 & & \tilde P_{\rm HI}\tilde P_{g{\rm HI}} & & \tilde P_{\rm HI}^2 
\end{pmatrix}\;. \label{e4.4}
\end{equation}
Here $\Delta k$ and $\Delta \mu$ are the bin-widths for $k$ and $\mu$, which we choose  following \citep{Karagiannis:2018jdt,Yankelevich:2018uaz}:
\begin{equation}
\Delta \mu = 0.04 \quad \text{and} \quad \Delta k = k_{\mathrm{f}}\;.\label{e4.5}
\end{equation}

In this paper our focus is on the improvements that the multi-tracer can deliver for pairs  (galaxy, HI IM) of high and very high redshift surveys. We assume that most of the standard cosmological parameters have been measured by {\em Planck} \cite{Planck:2018vyg}.
Since $f_{\mathrm{NL}}$ affects the amplitude and shape of the power spectrum on ultra-large scales, we include in the Fisher analysis the cosmological parameters $A_{s}$ and $n_s$ that directly affect these properties of the power spectrum. 

 {Furthermore, surveys with the tracers $A=g,$ HI are assumed each to have already established the redshift evolution $\alpha_A(z)$ of the Gaussian clustering biases, leaving only the amplitude parameters $b_{A0}$ in \eqref{e2.2} to be constrained (see \cite{Agarwal:2020lov} for justification of this assumption). 
We also assume that the average HI brightness temperature $\bar T_{\rm HI}(z)$  has been determined.
In summary, we assume that the {\em Planck} survey and the  single-tracer  measurements have already determined the standard cosmological and astrophysical quantities. 
We incorporate the uncertainties in the bias amplitude parameters and in the amplitude and slope of the primordial power spectrum, since our focus is on the local primordial non-Gaussianity and on parameters that are strongly linked to it. The multi-tracer approach is known to significantly reduce the impact of uncertainties in the clustering biases (e.g. \cite{Seljak:2008xr,McDonald:2008sh,Alonso:2015sfa,Fonseca:2015laa,Viljoen:2021ocx}).

These assumptions  lead to optimistic forecasts for $\sfnl$ -- but our aim is to find the {\em relative improvement from the multi-tracer}, rather than the precise values of $\sfnl$.}
Consequently, we consider the following cosmological and nuisance parameters:
\begin{equation}
\vartheta_{\alpha} = \big\{f_{\mathrm{NL}},\;A_{s},\;n_{s},\;b_{g0},\;b_{{\rm HI} 0} \big\}\,.\label{e4.3}
\end{equation}
The fiducial values for the LCDM cosmological parameters $A_{s},\;n_{s}$ are taken from 
{\em Planck} \cite{Planck:2018vyg} and we assume $\bar{f}_{\rm NL}=0$.
Derivatives in \eqref{e4.1} with respect to $f_{\mathrm{NL}}$, $A_{s}$  and $b_{A0}$ are computed analytically, while the $n_s$ derivative is computed using the 5-point stencil method, with step-length 0.1 \cite{Yahia-Cherif:2020knp}.

The Fisher forecast results, using the fiducial value of the non-Gaussian galaxy assembly bias parameter $p_g$ in \eqref{e2.9} and the fiducial values of the foreground-avoidance paramaters $k_{\rm fg}$ and $N_{\rm w}$ in \eqref{kfgflo} and \eqref{nwflo},  are shown in
\autoref{fig6} and \autoref{fig6a}, where we have marginalised over the bias nuisance parameters. These plots display the $1\sigma$  error contours for the multi-tracer pairs $A\otimes B={\rm H} \alpha\,\otimes$ H (Euclid-like\,$\otimes$\,HIRAX-like, high redshift) and LBG\,$\otimes$\,P (MegaMapper-like\,$\otimes$\,PUMA-like, very high redshift). 
The numerical values of $\sfnl$ are shown in \autoref{tab2}.

\begin{table}[!ht] 
\centering 
\caption{Marginalised  {68\% CL} errors on $f_{\mathrm{NL}}$  {from galaxy surveys $A=$ H$\alpha$ (Euclid-like), LBG (MegaMapper-like) and HI IM inteferometer-mode surveys $A=$ H (HIRAX-like) and P (PUMA-like). Both H and P have phases 1 and 2, with final number of dishes given in parentheses. Multi-tracer combinations are $A\otimes B= {\rm H}\alpha\otimes H$ (high redshift) and LBG\,$\otimes$\,P (very high redshift)}. The final column shows the percentage improvement in precision from the multi-tracer for each single tracer.
}\label{tab2} 
\vspace*{0.4cm}
\begin{tabular}{|l|l|c|} 
\hline 
Surveys $A,\, A\otimes B$ & 
{$\sigma(f_{\mathrm{NL}})$} 
& $\big[\sigma(A)-\sigma(A\otimes B)\big]/\sigma(A)$
\\ & & $\times 100\%$
\\ \hline\hline 

H$\alpha$        & 3.83      
& 21 ~(29)
\\
LBG              &  0.787    
& 28 ~(33)
\\
H\,256 ~(1024)     & 11.4 ~~(9.18)   
& 74 ~(70)
\\  
P\,5k ~~\,(32k)    & 1.47 ~~(1.31)   
& 61 ~(60)
\\  \hline

H$\alpha$ $\otimes$ H\,256 ~\,(1024) & 3.02~~~\,(2.71)  
&
\\ 
LBG $\otimes$ P\,5k ~(32k)           & 0.569~~(0.530)   
&
\\ \hline
\end{tabular}
\end{table}
~\\

 {\autoref{tab3} displays the changes in  $\sfnl$  when using the alternative estimates for $p_g$ in \eqref{e2.9} and the less optimistic values of $k_{\rm fg}$ and $N_{\rm w}$ in \eqref{kfgflo} and \eqref{nwflo}.}~\\
\begin{table}[h]  
\caption{ {As in \autoref{tab2} but showing the effect on the multi-tracer $\sfnl$ when: (a)~we change the non-Gaussian galaxy assembly bias parameter $p_{g}$ from the fiducial value $p_g=0.55$ to the alternative choices in \eqref{e2.9} and fix  $p_{\rm HI}=1.25$ as in \eqref{e2.8}; (b)~we enlarge the set of foreground avoidance parameters $(k_{\rm{fg}}, N_{\rm{w}})$ from the fiducial values $(0.005\,h{\rm Mpc}^{-1},\,1)$ to include  the less optimistic values $(0.01\,h{\rm Mpc}^{-1},\,3)$, and we consider all possible  pairs $(k_{\rm{fg}}, N_{\rm{w}})$.}} 
\centering 
\label{tab3} 
\vspace*{0.4cm}
\begin{tabular}{l c c  rr | rr} 
\hline\hline   
 $p_{g}$ & Surveys $ A\otimes B$ & $k_{\rm{fg}}[h{\rm Mpc}^{-1}]$  &\multicolumn{2}{c}{$N_{\rm{w}}=1$} &  \multicolumn{2}{c}{$N_{\rm{w}}=3$}
\\ [0.5ex]  
\hline   
 & &0.005 & 3.02 & (2.71) & 3.04 & (2.73)  \\[-0.5ex]
 & \raisebox{1.5ex}{H$\alpha$ $\otimes$ H\,256~(1024)} &0.01  & 3.23 & (2.95) & 3.24 & (2.96)  \\[-0.5ex]
\raisebox{3.0ex}{$0.55$} & &0.005  
& 0.569 & (0.530) & 0.596 & (0.560)  \\[-0.5ex]
& \raisebox{1.5ex}{LBG $\otimes$ P\,5k~(32k)} &0.01  & 0.613 & (0.568) & 0.635 & (0.594)  \\[1ex]\hline
& &0.005 & 5.36 & (5.14) & 5.38 & (5.17)  \\[-0.5ex]  
 & \raisebox{1.5ex}{H$\alpha$ $\otimes$ H\,256~(1024)}& 0.01  
&5.44 & (5.25) & 5.46 & (5.28)  \\[-0.5ex] 
\raisebox{3.0ex}{$1.0$} & & 0.005  
& 0.694 & (0.661) & 0.719 & (0.691)  \\[-0.5ex]
& \raisebox{1.5ex}{LBG $\otimes$ P\,5k~(32k)} &0.01  & 0.728 & (0.691) & 0.747 & (0.716)  \\[1ex]\hline
& &0.005 & 7.35 & (6.07) & 7.41 & (6.15) \\[-1ex]  
 & \raisebox{1.5ex}{H$\alpha$ $\otimes$ H\,256~(1024)}& 0.01  
&8.66 & (7.35) & 8.70 & (7.42) \\[1ex]  
\raisebox{1ex}{$1.5$} & & 0.005  
& 0.860 & (0.840) & 0.882 & (0.868)  \\[-0.5ex]
& \raisebox{1.5ex}{LBG $\otimes$ P\,5k~(32k)} &0.01  & 0.873 & (0.853) & 0.892 & (0.876)  \\[1ex]
\hline 
\end{tabular}  
\end{table}  
 

\section{Conclusion}\label{Conclusion}

 {We used  Fisher forecasting to estimate the precision on local primordial non-Gaussianity that is achievable by combining future galaxy and HI intensity mapping surveys (taking foregrounds and interferometer thermal noise into account).
We use a simplified model, incorporating uncertainties in the Gaussian clustering bias parameters and in the amplitude and slope of the primordial power spectrum, but fixing other parameters. Although this means that   our $\sfnl$ forecasts are optimistic, it allows us to estimate the relative improvement in precision from the multi-tracer.

The results of the Fisher analysis, shown in \autoref{fig6} and \autoref{fig6a} and summarised in \autoref{tab2}, confirm the effectiveness of using a multi-tracer combination of HI intensity mapping and spectroscopic galaxy samples, in order to enhance the precision in future measurements of local primordial non-Gaussianity. 
We have shown this for the first time in the case that the HI intensity mapping survey is performed in interferometer mode, as opposed to the single-dish mode used in previous forecasts.
For the fiducial choice of non-Gaussian galaxy assembly bias parameter $p_g$ and of intensity mapping foreground parameters $k_{\rm fg}, N_{\rm w}$, the relative improvement from the multi-tracer  over the best single-tracer constraint (from the galaxy samples) is $\sim 20-30\%$ (see \autoref{tab2}).

The multi-tracer also improves precision on $n_s$ and $A_s$, as shown in \autoref{fig6} and \autoref{fig6a}. It is noticeable that the HI intensity mapping interferometer-mode surveys provide stronger constraints on $n_s, A_s$ than the galaxy surveys, while the reverse is true for $\fnl$. The reason is that the $n_s, A_s$ signals are mainly dependent on intermediate scales, while the $\fnl$ signal is based on ultra-large scales. The HI interferometer-mode surveys provide better constraints on intermediate scales than the galaxy surveys. However, on ultra-large scales these surveys lose constraining power due to the foreground filters. 

Our assumption on the  {tracer}-dependent non-Gaussian galaxy assembly bias, i.e. that $b_{A\phi}\propto b_A-p_A$ as in \eqref{e2.8} and \eqref{e2.9}, also affects galaxies and HI intensity mapping differently. In the simplest universality model, $p_A=1$, there is no difference between galaxies and HI intensity in the form of the non-Gaussian galaxy assembly bias. In the fiducial case with $p_g<1$, it follows that $b_{g\phi}$ is larger than in the $p=1$ model. Consequently, the constraints for galaxy surveys on $\fnl$ are improved compared to $p=1$. By contrast, $p_{\rm HI}>1$ means that $b_{{\rm HI}\phi}$ is larger than in the $p=1$ model, so that $\fnl$ constraints are degraded for HI intensity mapping surveys compared to the $p=1$ model.

\begin{figure}[! ht]
\centering
\includegraphics[width=10cm]{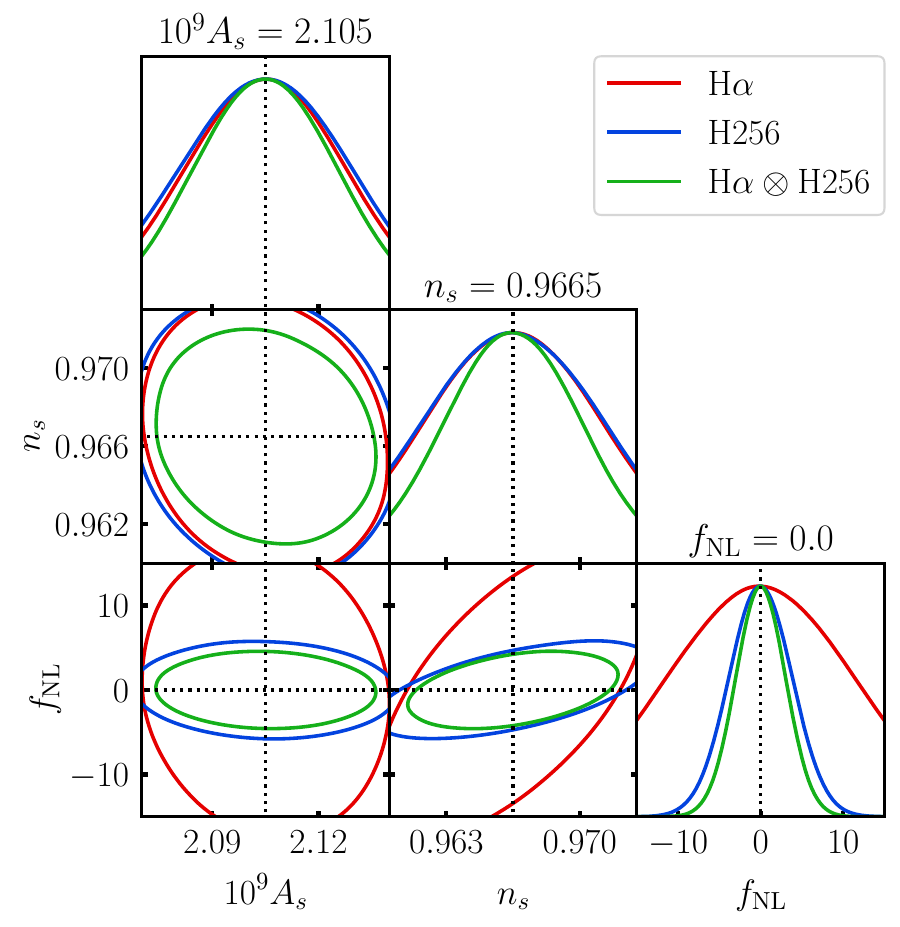}  
\includegraphics[width=10cm]{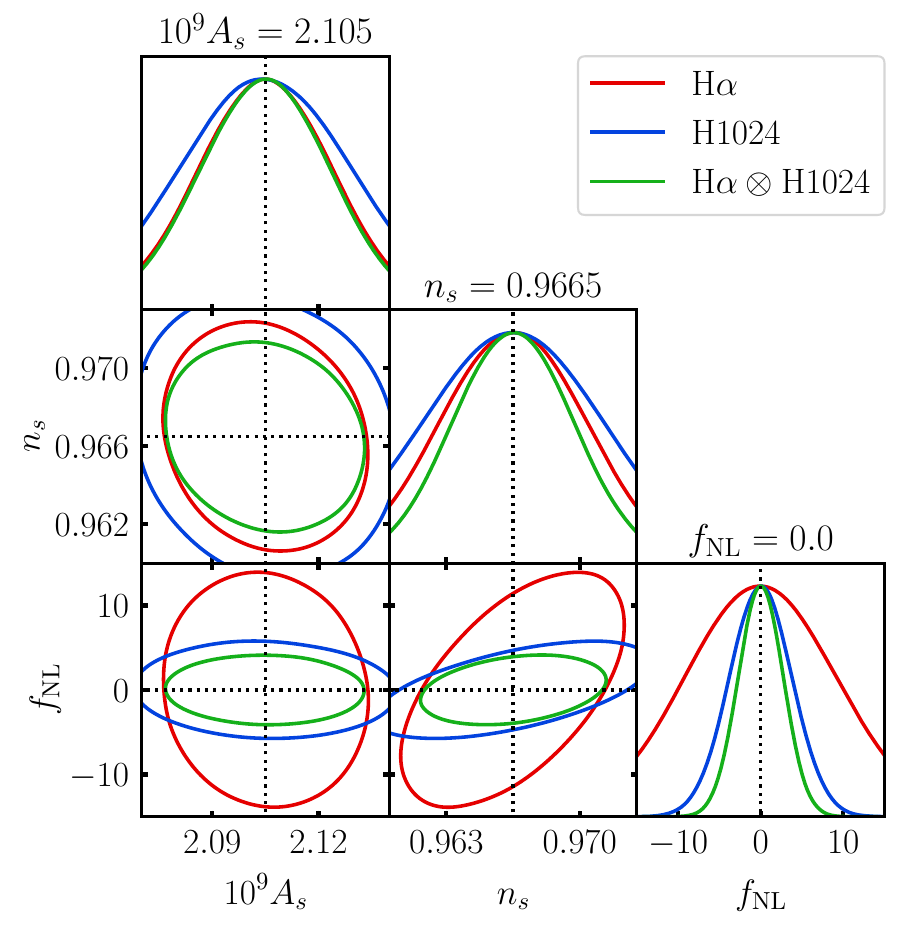}\\
\caption{ {Forecasted 1$\sigma$ (68\% CL) marginalised uncertainty contours for the local primordial non-Gaussianity parameter $\fnl$, together with the cosmological parameters $A_s$ and $n_s$, computed from the H$\alpha$ (Euclid-like) galaxy survey and the H (HIRAX-like) intensity mapping survey. {\em Top:} 256 H dishes; {\em bottom:} 1024 H dishes.}} \label{fig6}
\end{figure}

\begin{figure}[! ht]
\centering
\includegraphics[width=10cm]{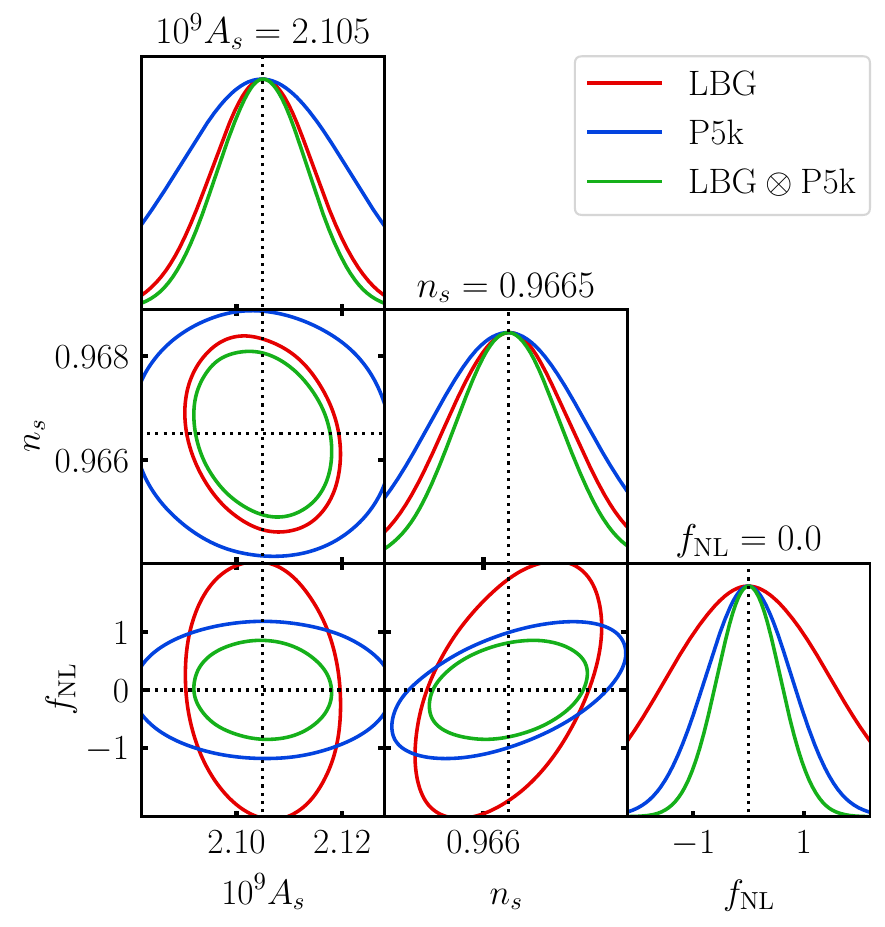}  
\includegraphics[width=10cm]{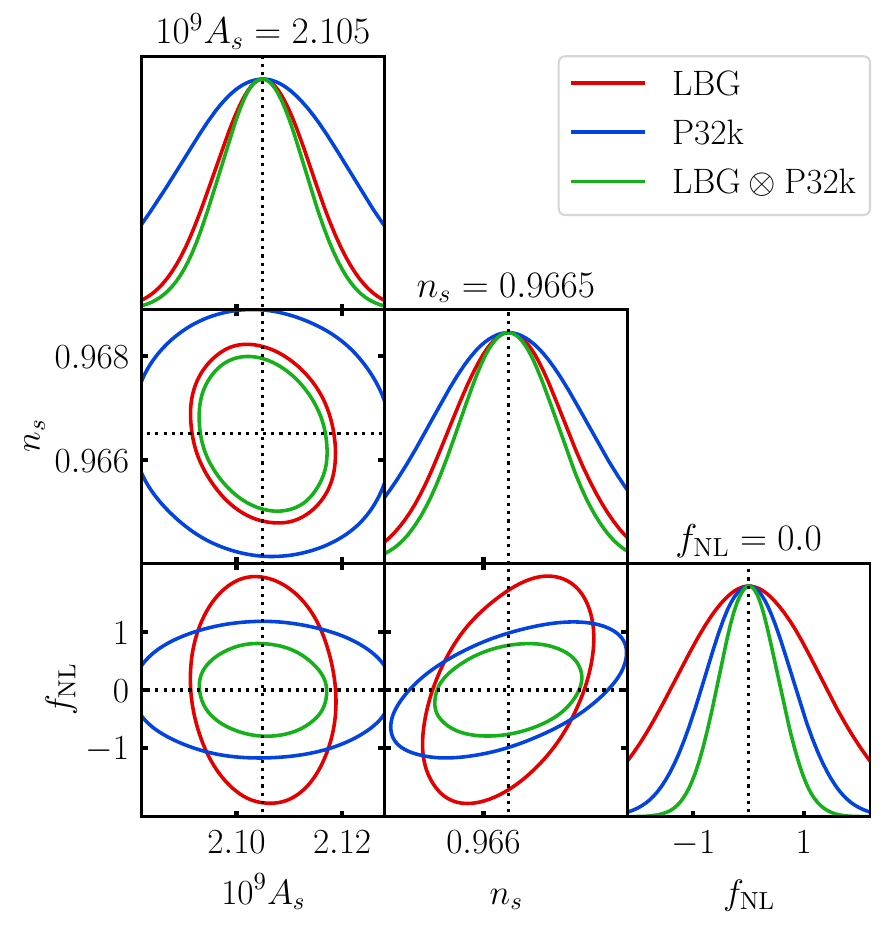}
\caption{ {As in \autoref{fig6} but for the LBG (MegaMapper-like) galaxy survey and the P (PUMA-like) intensity mapping survey. {\em Top:} 5000 P dishes; {\em bottom:} 32,000 P dishes.}} \label{fig6a}
\end{figure}
 
Single-tracer Fisher constraints for surveys similar to our mock surveys have been computed previously: e.g., \cite{Viljoen:2021ypp} for Euclid; \cite{Sailer:2021yzm} for MegaMapper and PUMA; \cite{Karagiannis:2019jjx} for HIRAX and PUMA. These constraints are qualitatively compatible with ours -- taking into account the following differences:
\begin{itemize}
\item Previous work uses the universality model for $b_{A\phi}$, while we use  \\eqref{e2.7}, \eqref{e2.8} and \eqref{e2.9}.
\item Our mock surveys have differences in the redshift range, sky area and observing time (for HI intensity mapping) compared to the surveys considered in previous work.
\item There are differences amongst the previous works, and with our paper, in the cosmological and nuisance parameters that are used and marginalised over.
\end{itemize}

\noindent To our knowledge, there are no previous works using the multi-tracer pairs that we analyse, so that comparisons of our multi-tracer results with previous results is not possible.

The best single-tracer and
multi-tracer $f_{\mathrm{NL}}$ precision is delivered at the highest redshifts, by LBG (MegaMapper-like) and LBG$\,\otimes\,$P (PUMA-like) (\autoref{fig6a}). The LBG survey on its own already achieves $\sigma(f_{\mathrm{NL}}) < 1$ (this is compatible with \cite{Sailer:2021yzm}), while P constraints (which are compatible with \cite{Karagiannis:2019jjx})  are weaker because of foreground contamination on ultra-large scales. 
When LBG and P are combined, we achieve an  improvement in $\sigma(f_{\mathrm{NL}})$ of  {$\sim 30\%$} over the single-tracer LBG.

 {Our fiducial foreground avoidance parameters are
fairly optimistic,} but we expect that further advances in foreground treatments will emerge. In particular:
\begin{itemize}
    \item 
density reconstruction methods, such as forward model reconstruction, are being developed to retrieve more long wavelength radial modes lost to  foregrounds (e.g \cite{Zhu:2016esh,Modi:2019hnu,Cunnington:2023jpq});
\item
advances in the stability and calibration of interferometers will reduce the impact of the wedge and in principle will be able in future to reduce its impact to negligible.
\end{itemize}

In addition, there are theoretical uncertainties in the non-Gaussian galaxy assembly bias parameter $p_g$. Our fiducial choice $p_g=0.55$ is based on galaxy samples selected by stellar mass, whereas the galaxy samples we use may have different $p_g$. We therefore assessed the impact on $\sfnl$ of different $p_g$ choices [see  \eqref{e2.9}], as well as less optimistic intensity mapping foreground-avoidance parameters [see  \eqref{kfgflo}, \eqref{nwflo}]. The results are summarised in \autoref{tab3}. It is apparent that the increase in $p_g$ leads to the largest increases in $\sfnl$. For example, if we consider the two best multi-tracer pairs, then  \autoref{tab3} shows that with the foreground parameters fixed at their fiducial values, the increase in the errors when changing $p_g$ from 0.55 to 1.5 is $124\%$ for H$\alpha\,\otimes\,$H\,1024 and 58\% for LBG$\,\otimes\,$P\,32k. This makes sense, since increasing $p_g$ reduces $b_{g\phi}$ and hence reduces the scale-dependent bias. Note that the degradation is significantly weaker in the best-performing multi-tracer LBG$\,\otimes\,$P\,32k. For the foreground parameters, the increase in  $N_{\rm w}$ to 3 primary beams has very little impact on $\sfnl$, while increasing the radial mode suppression parameter $k_{\rm fg}$ leads to a larger increase in $\sfnl$. In detail, increasing $N_{\rm w}$ from 1 to 3 leads to increases in $\sfnl$ of 
 $1\%$ for H$\alpha\,\otimes\,$H\,1024 and $6\%$ for LBG$\,\otimes\,$P\,32k.
 Changing $k_{\rm fg}$ from 0.005 to $0.01h$/Mpc leads to an increase in the errors of $9\%$  and $7\%$  respectively.

We conclude that the galaxy--HI multi-tracer constraints are not very sensitive to HI foreground avoidance parameters but are significantly degraded by increases in the non-Gaussian galaxy assembly bias parameter. 
Nevertheless, the improvements from the multi-tracer over the best single tracer is robust to the increases in $p_g$, $k_{\rm fg}$ and $N_{\rm w}$. The improvement ranges from 10 to 30\%.

Fisher forecasts are of course highly simplified compared to observational and data reality. In practice there are numerous systematics to be modelled and there are major computational constraints on the simulations necessary to build a data pipeline. Nevertheless, a simplified Fisher analysis has shown that multi-tracing galaxy and  HI intensity mapping interferometer-mode surveys at high to very high redshifts is worth exploring further. Furthermore, 
the multi-tracer will alleviate the impact of nuisance parameters and systematics in the galaxy and HI intensity mapping samples.  

Finally, we note that our forecasts use  a plane-parallel approximation in the Fourier power spectra. Future work will relax this assumption by including  wide-angle effects (see e.g. \cite{Viljoen:2021ocx,  Noorikuhani:2022bwc, Paul:2022xfx}).

\vfill
\noindent\textbf{Acknowledgements} \\
We thank Dionysios Karagiannis, Stefano Camera and Lazare Guedezounme for very helpful comments. We are supported by the South African Radio Astronomy Observatory (SARAO) and the National Research Foundation (Grant No. 75415). 
\clearpage

\bibliographystyle{JHEP}
\bibliography{reference_library}

\end{document}